\documentclass[11pt]{article}
\usepackage{amsthm,amsmath,epsf,amssymb,enumerate,array,color}
\usepackage[all,matrix,arrow,import,arc,poly,color,ps,dvips]{xy}

\newtheorem{thm}{Theorem}[section]        \newtheorem{lemma}[thm]{Lemma}	
\newtheorem{definition}[thm]{Definition}  	\newtheorem{prop}[thm]{Proposition}
  \newtheorem{conj}[thm]{Conjecture}	\theoremstyle{definition}

\DeclareFontFamily{U}{rsf}{} \DeclareFontShape{U}{rsf}{m}{n}{  <5> <6> rsfs5 <7> <8> <9> rsfs7 <10-> rsfs10}{}
\DeclareMathAlphabet\Scr{U}{rsf}{m}{n} \DeclareMathAlphabet\mathbi{U}{cmr}{bx}{it}

\def\CY{Calabi-Yau}	
\def\roof{\mbox{\tiny \mbox{$\!\vee$}}}	\def\comp{\mbox{\scriptsize \mbox{$\circ \,$}}}

\def\O{\mathcal{O}} \def\c#1{\mathcal{#1}}	\def\Coh{\mathfrak{Coh}}
\def\C{{\mathbb C}}\def\P{{\mathbb P}} \def\Q{{\mathbb Q}}\def\F{{\mathbb F}}
 \def\Z{{\mathbb Z}}
\def\D{\mathbf{D}}\def\K{\mathbf{K}}\def\L{\mathbf{L}} \def\R{\mathbf{R}}
\def\iso{\cong}
\def\H{\operatorname{H}}
    
\def\id{\operatorname{id}}
\def\Hom{\operatorname{Hom}} \def\sHom{\operatorname{\Scr{H}\!\!\textit{om}}}	
\def\Ext{\operatorname{Ext}}      \def\sExt{\operatorname{\Scr{E}\!\textit{xt}}}		\def\RHom{\R\!\operatorname{Hom}}

\def\Spec{\operatorname{Spec}}

\def\Pic{\operatorname{Pic}}		
	\def\SL{\operatorname{SL}}		
\def\SU{\operatorname{SU}}	\def\U{\operatorname{U{}}}	

\def\rk{\operatorname{rk}}	\def\dim{\operatorname{dim}}

\def\ch{\operatorname{\mathrm{ch}}}		\def\td{\operatorname{\mathrm{td}}}

\def\Ltensor{\mathbin{\overset{\mathbf L}\otimes}}

\def\ms#1{\mathsf{#1}}		\def\cal{\mathcal}
\def\poso#1{#1\save="x"!LD+<0pt,-0.5mm>;  "x"!RD+<0pt,-0.5mm>**\dir{.}\restore}
\def\Cone#1{\operatorname{Cone}\left( #1 \right)}
\def\ses#1#2#3{\xymatrix@1{0 \ar[r] & #1 \ar[r] & #2 \ar[r] & #3 \ar[r] & 0}}

\def\pplogo{\vbox{\kern-\headheight\kern -29pt
\halign{##&##\hfil\cr&{\ppnumber}\cr\rule{0pt}{2.5ex}&\ppdate\cr}}}
\makeatletter
\def\ps@firstpage{\ps@empty \def\@oddhead{\hss\pplogo}%
  \let\@evenhead\@oddhead 
}
\def\maketitle{\par
 \begingroup
 \def\thefootnote{\fnsymbol{footnote}}
 \def\@makefnmark{\hbox{$^{\@thefnmark}$\hss}}
 \if@twocolumn
 \twocolumn[\@maketitle]
 \else \newpage
 \global\@topnum\z@ \@maketitle \fi\thispagestyle{firstpage}\@thanks
 \endgroup
 \setcounter{footnote}{0}
 \let\maketitle\relax
 \let\@maketitle\relax
 \gdef\@thanks{}\gdef\@author{}\gdef\@title{}\let\thanks\relax}
\makeatother

\begin{document}
\setcounter{page}0
\def\ppnumber{\vbox{\baselineskip14pt
\hbox{hep-th/0510047}}}
\def\ppdate{October 2005} \date{}

\title{\bf \LARGE $\C^2/\Z_n$ Fractional branes and Monodromy		\\[10mm]}
\author{{\bf Robert L.~Karp} \thanks{karp@physics.rutgers.edu}		\\[2mm]
\normalsize  Department of Physics, Rutgers University \\
\normalsize Piscataway, NJ 08854-8019 USA				}

{\hfuzz=10cm\maketitle}

\vskip 1cm

\begin{abstract}
\normalsize
\noindent
We construct geometric representatives for the $\C^2/\Z_n$ fractional branes in terms of branes wrapping certain exceptional cycles of the resolution. In the process we use large radius and conifold-type monodromies, and also check some of the orbifold quantum symmetries. We find the explicit Seiberg-duality which connects our fractional branes to the ones given by the McKay correspondence. We also comment on the Harvey-Moore BPS algebras.
\end{abstract}

\vfil\break

\tableofcontents

\section{Introduction}    \label{s:intro}

Understanding the physical properties of D-branes  throughout the entire moduli space of a given Calabi-Yau compactification is an important and so far unsolved problem. Nevertheless much progress has been made in this direction. For $\c N=2$ Type II compactifications, $\pi$-stability and derived categories  seem to provide the most general framework so far, as it has been argued that topological B-branes are in one-to-one correspondence with the objects of the derived category of coherent sheaves on the \CY\ variety \cite{Douglas:2000gi}. For non-linear sigma-models on a large \CY\  this has been carefully checked \cite{Aspinwall:2001pu}.

Since the Kahler deformations are exact in the topological B-model, one expects the derived category description of topological B-branes to be valid at any point of the moduli space. There is ample evidence for this by now. $\c N=2$ Type II compactifications generically have a rich phase structure. The description of B-branes in the various phases is quite different, and the expected equivalence gives rise to interesting mathematical statements. The best known example of this sort is the celebrated McKay correspondence.

This shifted the question from asking what the D-branes are at a given point in moduli space to asking which ones are stable, more precisely $\pi$-stable, and therefore physical. 

Determining the set of stable branes is cumbersome. The most workable method suggests to start at a point where one has a good understanding of stability, e.g., a large radius point, where $\pi$-stability reduces to $\mu$-stability, and try to catalog what objects are lost and gained as the Kahler moduli are varied ``adiabatically'' \cite{Aspinwall:2001dz}.  

Orbifolds provide a rich testing ground for these ideas. In this case D-branes  can be described explicitly as boundary states in a solvable conformal field theory (CFT). But there is another description, originating from the world-volume theory of the D-brane probing the orbifold singularity, which is a quiver gauge theory \cite{Douglas:Moore}. In this language D-branes  are objects in the derived category of representations of the  quiver. The McKay correspondence gives an equivalence between this category and the derived category of coherent sheaves on the resolved space \cite{McKay,Mukai:McKay}.

As we said earlier, the McKay correspondence is a prototype of what happens in general: in different patches of the moduli space one has very different looking descriptions for the D-branes, which sit in {\em inequivalent} categories, but if one passes to the derived category then they become equivalent. Therefore it makes sense to talk about a geometric representation for a brane at any point in moduli space. Passing from an abelian  category to the derived category is physically motivated by brane--anti-brane annihilation, thorough  tachyon condensation \cite{Douglas:2000gi,Witten:1998cd}. 

In the quiver representation language one can consider the simple representations, i.e., those that have no non-trivial subrepresentations. These correspond to  fractional branes \cite{Diaconescu:wrap}. Their physical interpretation using  $\pi$-stability is quite simple: at the orbifold point the space-filling D3-brane becomes marginally stable against decaying into the fractional branes. In fact this phenomenon is more general than orbifolds, and should apply to a D3-brane at any \CY\ singularity. This has been understood in great detail for the conifold \cite{Aspinwall:2002ke}.

Although the  fractional branes are obvious in the quiver representation language, their geometric interpretation is quite unclear. The McKay correspondence tells us that there should be objects (bundles or perhaps complexes) on the resolved space whose $\Ext^1$-quiver is the one we started with. One of the central problems in this area is to find these objects. 

As a warm-up exercise one can try to determine the K-theory class of the fractional branes. So far, even this question has been answered only in a limited context, using mirror symmetry techniques \cite{Diaconescu:1999dt} or the McKay correspondence\footnote{ For the ample physics literature on this subject see, e.g., \cite{Paul:TASI2003} and references therein.}. 

Our general goal is to get a deeper understanding of the geometry of  fractional branes, going beyond K-theory. And we will do this without resorting to mirror symmetry or the McKay correspondence. Instead we use the quantum symmetry of an orbifold theory to generate the fractional branes as an orbit. 

In the present paper we investigate several collections of  fractional branes for the $\C^2/\Z_n$ orbifolds, while the case of $\C^n/G$ will be investigated elsewhere \cite{en:fracC3}. Ultimately one would like to understand the world-volume theory of D-branes at an arbitrary point in the moduli space of a compact \CY . Studying examples where one has at his disposal different methods hopefully will teach us the ``mechanics'' of the geometric approach. We hope to return to some phenomenologically more interesting examples, like \cite{Berenstein:2005xa} and \cite{Verlinde:2005jr}, in the future.

As a byproduct of the techniques developed in this paper, by the end, we will have performed some very strong consistency checks of the ``D-brane derived category'' picture. As we will see shortly, the functors implementing the monodromy transformations are not simple by any measure. The relations they satisfy would be very hard to guess without physical input.

In the last part of the paper we use stacky methods to provide several collections of fractional branes. This relies on an extension of the McKay correspondence due to Kawamata. These branes are naturally associated to regions of the moduli space where we have no solvable CFT description, or a reliable supergravity approximation either. Fortunately, the algebro-geometric tools are powerful enough to deduce what we want.

The organization of the paper is as follows. In Section~2 we use toric methods to investigate the geometry of the $\C^2/\Z_3$ model together with its Kahler moduli space.  Section~3 starts with a review of the Fourier-Mukai technology, and then it is applied to the various  $\C^2/\Z_3$ monodromies. Using the detailed structure of the moduli space we also prove the $\Z_2$ quantum symmetry at the $\Z_2$ point. To our knowledge, this is the first proof of this sort. In  Section~4 we use the $\Z_3$ monodromy to produce a collection of fractional branes. In  Section~5 this collection is compared to the one given by the McKay correspondence. This produces an interesting Seiberg duality, which allows us to extend our result to  $\C^2/\Z_n$  in general. Then we turn to a collection of fractional branes on the partially resolved $\C^2/\Z_n$ orbifold, with the use of a generalization of the McKay correspondence by Kawamata. The partially resolved orbifold is singular, and it is not a global quotient either,  hence it is particularly pleasing that  we can handle it directly by geometric methods. We conclude with some thoughts on the algebra of BPS states and the superpotential. The appendixes contain some spectral sequences that are used endlessly throughout the paper.

\section{{$\C^2/\Z_3$}{} geometries}\label{s:c2z3}

In this section we review some aspects of the $\C^2/\Z_3$ geometric orbifold and the associated CFT. First we work out the relevant toric geometry of $\C^2/\Z_3$, then we turn our attention to the moduli space of complexified Kahler forms, and in particular its discriminant loci. We pay particular attention to the singularities in the moduli space. Most of the material in this section must be known to the experts, but we could not find suitable references for it.

\subsection{The toric geometry of {$\C^2/\Z_3$}{}}   

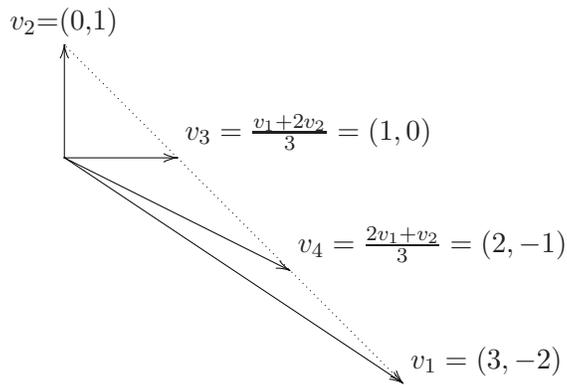
\begin{figure}[h]
\begin{equation}\nonumber
\begin{xy} <1.5cm,0cm>:
{\ar 0;(0,1) }, (0,1.2)*\txt{$v_2$=(0,1)}
,{\ar (0,0);(1,0) *+!LD{v_3=\frac{v_1+2v_2}{3}=(1,0)}}
,{\ar 0;(2,-1) *+!LD{v_4=\frac{2v_1+v_2}{3}=(2,-1)}}
,{\ar 0;(3,-2) *+!LD{v_1=(3,-2)}}
,{\ar@{-}@{.>} (0,1);(3,-2) }
\end{xy}
\end{equation}
  \caption{The toric fan for the resolution of the $\C^2/\Z_3$ singularity.}
  \label{f:fan}
\end{figure}

The $\C^2/\Z_3$ variety with the supersymmetric $\Z_3$ action 
\begin{equation}
(z_1,z_2)\mapsto (\omega z_1,\omega^2 z_2)\,, \qquad \omega ^3=1\,,
\end{equation}
is toric, and a convenient representation for it is provided by the fan in Fig.~\ref{f:fan}. 
More precisely, the $\C^2/\Z_3$ variety consists of only one cone, generated by the vertices $v_1$ and $v_2$. In this figure we also included the divisors corresponding to the crepant resolution of the singularity. We denote the resolved space by $X$. The exceptional locus of the blow-up consists of the divisors corresponding to the rays $v_3$ and $v_4$. Let us denote the divisors associated to $v_i$ by $D_i$. As we will see shortly, in Eq.~(\ref{e2}) below, $D_3$ and $D_4$ are both $-2$ curves.

There are two linear equivalence relations among the divisors:
\begin{equation} \label{se1}
3D_1+D_3+2D_4 \sim 0 \,, \quad -2D_1+D_2-D_4 \sim 0\,.
\end{equation}
From the toric diagram in Fig.~\ref{f:fan}  one can immediately read out the intersection relations:\footnote{Since $\C^2/\Z_3$ is non-compact, one restricts to the intersections of the compact cycles, in this case $D_3$ and $D_4$.}
\begin{equation}
D_2D_3=D_1D_4=1 \, ,\quad D_3D_4=1\,.
\end{equation}
Using the linear equivalences (\ref{se1}) we obtain the expected results
\begin{equation}\label{e2}
D_3^2=D_4^2=-2\,, \quad D_2D_4=D_1D_3=0\,.
\end{equation}

From the geometry it is also clear that the curves $D_3$ and $D_4$ are the generators of the Mori cone of effective curves. By Lemma 3.3.2 in \cite{Cox:Katz} there is a bijection between the Mori cone generators and the generators of the  lattice of relations of the point-set $\c A=\{v_1,\ldots, v_4 \}$. In particular, $D_3$ and $D_4$ yield the relations\footnote{In the gauged linear sigma model language the rows of this matrix represent the $\U(1)$ charges of the chiral superfields. }
\begin{equation}\label{e:chargem}
Q =
\left(\!\!
\begin{array}{rrrr}
0&1&-2&1\\
1&0&1&-2
\end{array}
\right)\,.
\end{equation}

The Kahler cone is dual to the Mori cone, and in our case both are two dimensional. It is immediate from the intersection products in (\ref{e2}) that the ordered pair $\{D_2,D_1\}$ is dual to the ordered pair $\{D_3,D_4\}$. The precise ordering will play an important role in what follows.

Since we are in two complex dimensions, an irreducible divisor is a curve. This leads to potential confusion. To avoid it, we refer to the {\em curves} represented by the divisors $D_i$ as $C_i$, and these live in the second homology $\H_2(X,\Z)$, while $D_i$ will refer to their  Poincare duals, which  live in the second cohomology $\H^2(X,\Z)$. In this notation we can rephrase the earlier result:
 \begin{equation}\label{e:ce1}
\{D_2,D_1\}\in\H^2(X,\Z)\quad \mbox{\rm is dual to}\quad \{C_3,C_4\}\in\H_2(X,\Z)\,.
\end{equation}

\subsection{The {$\C^2/\Z_3$}{} moduli space}   

\begin{figure}
\begin{equation}\nonumber
\begin{xy} <4cm,0cm>:
{\ar (0,0);(.5,0) *+!L{(1,0)\,x_3}}
,{\ar 0;(0,.6) *+!U{(0,1)\,x_4}}
,{\ar@{-} 0;(.5,-.8) *+!LU{(1,-2)}}
,{\ar@{-} 0;(-.8,.5) *+!R{(-2,1)}}
,(.8,.6)*+[F]\txt{Smooth\\ \ phase},
,(.3,.3)*+[o][F.]{{\mathcal C}_1}
,(-.5,-.6)*+[F]\txt{$\Z_3$  phase}
,(-.3,-.3)*+[o][F.]{{\mathcal C}_2}
,(.8,-.6)*+[F]\txt{$\Z_2$  phase}
,(.4,-.3)*+[o][F.]{{\mathcal C}_3}
,(-.5,.6)*+[F]\txt{$\Z_2$  phase}
,(-.2,.3)*+[o][F.]{{\mathcal C}_4}
\end{xy}
\end{equation}
  \caption{The phase structure of the $\C^2/\Z_3$ model.}
  \label{fig:Z3}
\end{figure}
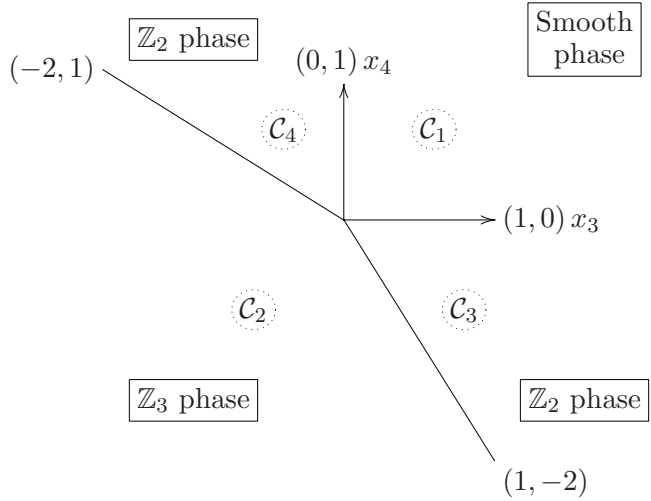

The point-set $\c A=\{v_1,\ldots, v_4 \}$ admits four obvious triangulations. Therefore in the language of \cite{Witten:GLSM,Aspinwall:1994nu}  the gauged linear sigma model has four phases. The secondary fan has its rays given by the columns of the matrix (\ref{e:chargem}), and is depicted in Fig.~\ref{fig:Z3}. The four phases are as follows: the completely resolved smooth phase; the two phases where one of the $\P^1$'s has been blown up to partially resolve the $\Z_3$ fixed point to a $\Z_2$ fixed point; and finally the $\Z_3$ orbifold phase. The $\Z_2$ phases corresponding to the cones ${\mathcal C}_3$ and ${\mathcal C}_4$ can be reached from the smooth phase ${\mathcal C}_1$ by blowing down the curves $C_3$ resp. $C_4$.

The discriminant locus of singular CFT's can be computed using the Horn parametrization \cite{GKZ:book,MP:SumInst}. We briefly review the general construction.
Let us denote by $Q=(Q_i^a)$, ${i=1,\ldots,n}$, ${a=1,\ldots,k}$, the matrix of charges appearing in Cox's holomorphic quotient construction of a toric variety \cite{Cox:HoloQout}. The primary component of the discriminant, $\Delta_0$, is a rational variety\footnote{Meaning that it is birational to a projective space.}. 
Horn uniformization gives an explicit rational parametrization for $\Delta_0$. Accordingly, we introduce $k$ auxiliary variables, $s_1,\ldots,s_k$ and form the linear combinations
\begin{equation}\label{e:hu1}
\xi_i=\sum_{a=1}^k Q_i^a\,s_a\,,\qquad \mbox{for all $i=1,\ldots,n$}\,.
\end{equation}
Let $x_a$, ${a=1,\ldots,k}$ be local coordinates on the moduli space of complex structures of the mirror.
$\Delta_0$ then has the following parameterization:
\begin{equation}\label{e:hu2}
x_a=\prod_{i=1}^n \xi_i^{\;Q_i^a}\,,\qquad \mbox{for all $a=1,\ldots,k$}\,.
\end{equation}

In our context the matrix of charges in question is $Q$ from Eq.~(\ref{e:chargem}). Let $x_3$ and $x_4$ be the local coordinates on the moduli space of complex structures of the mirror. One can use the mirror map so that $x_3$ and $x_4$ give coordinates on the Kahler moduli space of $X$ as well. Applying the Horn uniformization equations (\ref{e:hu1}) and (\ref{e:hu2}) gives 
\begin{equation}\label{e:cpxc}
x_3=\frac{s_1(s_1-2s_2)}{(2s_1-s_2)^2}\,, \qquad x_4=-\frac{s_2(2s_1-s_2)}{(s_1-2s_2)^2}\,.
\end{equation}
Since in our case $\sum_{i=1}^{4} Q_i^a=0$, for both $a=1,2$ the above equations are homogeneous, and therefore $x_3$ and $x_4$ depends only on the ratio $s_1/s_2$. Eliminating $s_1/s_2$ gives the sought after equation for $\Delta_0$:
\begin{equation}\label{eq:dZ3}
\Delta_0 = 27x_3^2x_4^2-18x_3x_4+4x_3+4x_4-1\,.
\end{equation}
In fact this is the only component of the discriminant.

The discriminant curve itself is singular. It has a unique singular point at $(x_3,x_4)=(1/3,1/3)$. To see the nature of the singularity we choose convenient coordinates around the singularity
\begin{equation}
x_3 = \frac{1}{3}+ (y_1+y_2)\,,\quad x_4=\frac{1}{3}+(y_1-y_2)\,,
\end{equation}
in terms of which the leading terms of $\Delta_0$ are\footnote{We discarded an overall factor of 12.}
\begin{equation}
3(y_1^3-y_1 y_2^2)+y_2^2\,.
\end{equation}
Since the $y_1 y_2^2$ term is subleading compared to the other two, $\Delta_0$ has a cusp at $(x_3,x_4)=(1/3,1/3)$.\footnote{A similar fact has been noted in \cite{navigation}.} This will be important later on, in Section \ref{s:M22}, since it allows one to have different monodromies around different parts of the primary component of the discriminant.

The orbifold points in the moduli space are themselves singular points. This fact is related to the quantum symmetry of an orbifold theory. For either of the $\Z_2$ points, the homogenous coordinate ring construction of Cox \cite{Cox:HoloQout} shows a $\C^2/\Z_2$ singularity with weights $(1,-1)$. Alternatively, using the ``old'' -- group algebra $\C[\sigma^{\roof}]$ of the dual cone $\sigma^{\roof}$ --  construction \cite{Fulton:T}, one arrives at the affine scheme $\Spec \C[y^{-1},x^2y,x] = \Spec \C[u,v,w]/(u v-w^2)$. At the $\Z_3$ point the moduli space locally is of the form $\C^2/\Z_3$, with weights $(1,2)$.

\section{ {$\C^2/\Z_3$}{} monodromies}    \label{s:m}

We start this section with a brief review of Fourier-Mukai functors. Then we  express the various  monodromy actions on D-branes in terms of Fourier-Mukai equivalences. The remaining part of the section deals with rigorously proving an identity which is the analog in the Fourier-Mukai language of the $\Z_2$ quantum symmetry at a partially resolved point in the moduli space.

\subsection{Fourier-Mukai functors}\label{s:fmf1}

For the convenience of the reader we review some of the key notions concerning Fourier-Mukai functors, and at same time specify the conventions used. We will make extensive use of this technology in the rest of the paper.  Our notation follows \cite{Horj:EZ}.  

Given two non-singular proper algebraic varieties, $X_1$ and $X_2$, an object ${\cal K} \in \D(X_1\! \times\! X_2)$ determines a functor of triangulated categories $\Phi_{\cal K}\!: \D(X_1) \to \D(X_2)$ by the formula\footnote{
$\D(X)$ denotes the bounded derived category of coherent sheaves on $X$. $\R p_{2*}$ is the total right derived functor of $p_{2*}$, i.e., it is an exact functor from $\D(X)$ to $\D(X)$. Similarly, $\Ltensor$ is the total left derived functor of $\otimes$. In later sections these decorations will be subsumed.}
\begin{equation}
\Phi_{\cal K}(A):=\R p_{2*} \big(\,{\cal K} \Ltensor p_1^*(A)\,\big)\,,
\end{equation}
where $p_i\!: X\! \times\! X \to X$ is projection to the $i$th factor:
\begin{equation}
\xymatrix{
  &X_1 \!\times\! X_2\ar[dl]_{p_1}\ar[dr]^{p_2}&\\
  X_1 & & X_2\,.}
\end{equation}
The object ${\cal K} \in \D(X_1\!\times\! X_2)$ is called the {\bf kernel} of the Fourier-Mukai functor $\Phi_{\cal K}$.

It is convenient to introduce the {\bf external tensor product} of two objects $A\in\D(X_1)$ and $B\in\D(X_2)$ by the formula
\begin{equation}
A\boxtimes B=p_2^*A\Ltensor p_1^*B\,.
\end{equation}

The importance of Fourier-Mukai functors when dealing with derived categories stems from the following theorem of Orlov:\footnote{Theorem 2.18 in \cite{Orlov:96}. The theorem has been generalized for smooth quotient stacks associated to normal projective varieties \cite{Kawamata:DC}.}
\begin{thm} \label{thm:orlov}
Let $X_1$ and $X_2$ be smooth projective varieties.
Suppose that $\mathsf{F}\!: \D(X_1)\to\D(X_2)$ is an equivalence of triangulated categories. Then there exists an object $\c K\in \D(X_1\!\times \! X_2)$, unique up to isomorphism, such that the functors $\mathsf{F}$ and $\Phi_{\c K}$ are isomorphic.
\end{thm}

The first question to ask is how to compose Fourier-Mukai (FM) functors. Accordingly, let $X_1$ $X_2$ and $X_3$ be three non-singular varieties, while let ${\cal F} \in \D(X_1\! \times\! X_2)$ and ${\cal G} \in \D(X_2\! \times\! X_3 )$ be two kernels. Let $p_{i j}\colon X_1\! \times\! X_2\! \times\! X_3\to X_i\! \times\! X_j$ be the projection map. A well-known fact is the following:
\begin{prop}\label{prop1}
The composition of the functors $\Phi_{\cal F}$ and $\Phi_{\cal G}$  is given by the formula
\begin{equation}
\Phi_{\cal G}\comp \Phi_{\cal F} \simeq\Phi_{\cal H}\,,\quad {\rm where}\quad
{\cal H}=\R p_{13*} \big(\, p_{23}^* ( {\cal G})\Ltensor  p_{12}^* ({\cal F})\big)\,.
\end{equation}
\end{prop}

Prop.~\ref{prop1} shows that  composing two FM functors gives another FM functor, with a simple kernel.
The composition of the kernels ${\cal  F}$ and ${\cal G} \in \D(X\times X)$ is therefore defined as
\begin{equation} \label{def:comp}
{\cal G}\star {\cal F} := \R p_{13*}\big(\, p_{23}^* ({\cal G})\Ltensor  p_{12}^* ({\cal F})\, \big).
\end{equation}

There is an identity element for the composition of kernels:
$\delta_* (\O_X),$ where $\delta : X \hookrightarrow X\! \times\! X$
is the diagonal embedding. For brevity we will denote $\delta_* (\O_X)$ by $\O_\Delta$:
\begin{equation}
\O_\Delta:=\delta_* (\O_X)\,.
\end{equation}
$\O_\Delta=\delta_* (\O_X)$ has the expected properties:
\begin{equation}\label{fm:id}
\O_\Delta\star{\cal G }={\cal G }\,\star\O_\Delta={\cal G }\,, \quad \mbox{for all ${\cal G} \in \D(X\times X)$}.
\end{equation}

Finally, the functors
\begin{equation}\label{trig:mor}
\begin{split}
&\Phi_{23}\!: \D(X_1\! \times\! X_2)\to \D(X_1\! \times\! X_3),\quad
 {\cal G }_{23}\in\D(X_2\! \times\! X_3),\quad
  \Phi_{23} (-):= {\cal G }_{23} \star -\,, \\
&\Phi_{12}\!: \D(X_2\! \times\! X_3)\to \D(X_1\! \times\! X_3),\quad
 {\cal G }_{12}\in\D(X_1\! \times\! X_2),\quad
  \Phi_{12} (-):= - \star {\cal G }_{12}\,,
\end{split}
\end{equation}
are morphisms between triangulated categories, i.e., they {\em preserve} distinguished triangles.

The composition of kernels is also associative
\begin{equation}\label{eq:assoc}
{\cal G}_{3} \star ({\cal G}_{2} \star
{\cal G}_{1}) \cong ({\cal G}_{3} \star {\cal G}_{2}) \star {\cal G}_{1}\,.
\end{equation}

Now we have all the technical tools ready to study the monodromy actions of physical interest.

\subsection{Monodromies in general}

The moduli space of CFT's contains the moduli space of Ricci-flat Kahler metrics. This, in turn, at least locally has a product structure, with the moduli space of Kahler forms being one of the factors. This is the moduli space of interest to us. In what follows we study the physics of D-branes as we move in the moduli space of complexified Kahler forms. This space is a priori non-compact, and its compactification consists of two different types of boundary divisors. First we have the {\em large volume} divisors.  These correspond to certain cycles being given infinite volume. The second type of boundary divisors are the irreducible components of the {\em discriminant}. In this case the CFT becomes singular. Generically this happens because some D-brane (or several of them, even infinitely many) becomes massless at that point, and therefore the effective CFT description breaks down. For the quintic this breakdown happens at the well known conifold point \cite{Strominger:1995cz}.

The monodromy actions around the above divisors are well understood. We will need a more abstract version of this story, in terms of Fourier-Mukai functors acting on the derived category, which we now recall.\footnote{For an extensive treatment of monodromies in terms of Fourier-Mukai functors see \cite{en:Horja,Distler:DC}.} 

Large volume monodromies are shifts in the $B$ field: ``$B\mapsto B+1$''. If the Kahler cone is higher dimensional, then we need to be more precise, and specify a two-form, or equivalently a divisor $D$. Then the monodromy becomes $B\mapsto B+D$. We will have more to say about the specific $D$'s soon. 

The simplest physical effect of this monodromy on a D-brane is to shift its charge, and this translates in the Chan-Paton language into tensoring with the line bundle $\O_X(D)$. This observation readily extends to the derived category:
 \begin{prop}\label{p:lr} 
The large radius monodromy associated to the divisor $D$ is
\begin{equation}
   \ms{L}_{D}(\mathsf{B}) = \mathsf{B}\Ltensor \O_X(D)\,,\qquad \mbox{for all $\mathsf{B} \in \D(X)$}\,.
\end{equation}
Furthermore, this is a Fourier-Mukai functor $\Phi_{{\cal L}}$, with kernel 
\begin{equation}
{\cal L}=\delta_*\O_X(D)\,,
\end{equation}
where  $\delta \!: X \hookrightarrow X\! \times\! X$ is again the diagonal embedding.
\end{prop}
Since we haven't found a suitable reference for this statement, we are going to include its proof. This will also serve as a warm-up exercise in the techniques that we use later on. From now on for brevity we are going to {\em suppress} most decorations $\L$, $\R$ and $ \Ltensor$ from the derived functors during computations.
\begin{proof}
All we need to show is that the Fourier-Mukai functor $\Phi_{\delta_*\O_X(D)}$ has the desired action. By its definition
\begin{equation}
\Phi_{\delta_*\O_X(D)}(\mathsf{B})=p_{2*}\big(\,\delta_*\O_X(D) \Ltensor  p_1^*(\mathsf{B})\big).
\end{equation}
Using the projection formula gives
\begin{equation}
\Phi_{\delta_*\O_X(D)}(\mathsf{B})=p_{2*}\delta_*\big(\O_X(D) \Ltensor  \delta^*p_1^*(\mathsf{B})\big).
\end{equation}
But $p_{2*}\,\delta_*=(p_2\comp\delta)_*=1_{\!X*}$ and $\delta^*p_1^*=(p_1\comp \delta)^*=1_{\!X}^{\;*}$, and this completes the proof.
\end{proof}

Now we turn our attention to the conifold-type monodromies. We will need the following conjecture from \cite{en:Horja}:\footnote{The conjecture goes back to Kontsevich, Morrison and Horja. We refer to \cite{en:Horja} for more details.} 
\begin{conj}\label{conj:a}
If we loop around a component of the discriminant locus associated with a single D-brane $\mathsf{A}$ (and its translates) becoming
massless, then this results in a relabeling of D-branes given by the autoequivalence of the derived category $\D(X)$
\begin{equation}
\mathsf{B}\longmapsto
\Cone{\RHom_{\D(X)}(\mathsf{A},\mathsf{B})\Ltensor\mathsf{A}\longrightarrow \mathsf{B}}\,.
\end{equation}
\end{conj}
 
This action is again  of Fourier-Mukai type. Lemma 3.2 of \cite{ST:braid} provides us with the following simple relation for any $\mathsf{B}\in\D(X)$:
\begin{equation}
\Phi_{ \Cone{ \mathsf{A}^{\roof}\boxtimes\,\mathsf{A} \to \O_\Delta} }(\mathsf{B}) \iso
\Cone{ \R\!\Hom_{\D(X)}(\mathsf{A},\mathsf{B})\Ltensor\mathsf{A}\longrightarrow \mathsf{B} }\,,
\end{equation}
where for an object $\mathsf{A}\in\D(X)$ its dual by definition is
\begin{equation}
\mathsf{A}^{\roof}= \R\!\!\sHom_{\D(X)}( \mathsf{A}, \O_X)\,.
\end{equation}

Since the functor $\Phi_{ \Cone{ \mathsf{A}^{\roof}\boxtimes\,\mathsf{A}\to\O_\Delta } }$ will play a crucial role, we introduce a special notation for it:
\begin{equation}\label{e:refl}
\ms{T}_{\mathsf{A}} :=
\Phi_{ \Cone{\mathsf{A}^{\roof}\boxtimes\,\mathsf{A}\to\O_\Delta } }\,,\quad
\ms{T}_{\mathsf{A}}(\mathsf{B}) =
\Cone{ \R\!\Hom_{\D(X)}(\mathsf{A},\mathsf{B})\Ltensor\mathsf{A}\longrightarrow \mathsf{B} }\,.
\end{equation}

The question of when is $\ms{T}_{\mathsf{A}}=
\Phi_{\Cone{ \mathsf{A}^{\roof}\boxtimes\,\mathsf{A}\to\O_\Delta } }$ an autoequivalence has a simple answer. For this we need the following definition:
\begin{definition} \label{def:spherical}
Let $X$ be smooth projective \CY\ variety of dimension $n$. An object $\mathsf{E}$ in $\D(X)$ is called {\em n-spherical} if $\Ext^r_{\D(X)}(\mathsf{E},\, \mathsf{E})$ is equal to $\H^r(S^n,\,\C)$, that is $\C$ for $r = 0,n$ and zero in all other degrees.
\end{definition}

One of the main results of \cite{ST:braid} is the following theorem:
\begin{thm}({\rm Prop. 2.10 in \cite{ST:braid})}
If the object $\mathsf{E}\in \D^b(X)$ is n-spherical then the functor $\ms{T}_{\mathsf{E}}$ is an autoequivalence.
\end{thm}

This brief review brings us to a point where we can apply this abstract machinery to study the $\C^2/\Z_3$ monodromies, and eventually use them to construct the fractional branes.

\subsection{{$\C^2/\Z_3$}{} monodromies}	\label{s:loops}

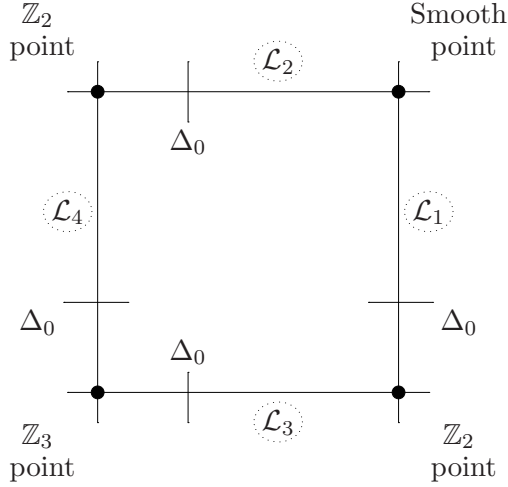
\begin{figure}
\begin{equation}\nonumber
\begin{xy} <4cm,0cm>:
{\ar@{-} (-.1,0);(1.1,0) }	,{\ar@{-} (1,.1);(1,-1.1) }	,{\ar@{-} (1.1,-1);(-.1,-1)}	,{\ar@{-} (0,.1);(0,-1.1)} 
,(1.2,.2)*\txt{Smooth\\ \ point}
,(.6,.1)*+[o][F.]{{\mathcal L}_2}
,(1.2,-1.2)*\txt{$\Z_2$\\ \ point}
,(1.1,-.4)*+[o][F.]{{\mathcal L}_1}
,(-.2,-1.2)*\txt{$\Z_3$\\ \ point}
,(.6,-1.1)*+[o][F.]{{\mathcal L}_3}
,(-.2,.2)*\txt{$\Z_2$\\ \ point}
,(-.1,-.4)*+[o][F.]{{\mathcal L}_4}
,(0,0)*+=[o]=<1.7mm>[F**:black][black]{.},(0,-1)*+=[o]=<1.7mm>[F**:black][black]{.}
,(1,0)*+=[o]=<1.7mm>[F**:black][black]{.},(1,-1)*+=[o]=<1.7mm>[F**:black][black]{.}
,{\ar@{-} (0.3,0.1);(0.3,-.1)   *+!U{\Delta_0}}		,{\ar@{-} (0.3,-1.1);(0.3,-.8)   *+!U{\Delta_0}}			
,{\ar@{-} (0.9,-.7);(1.2,-.7)   *+!U{\Delta_0}}		,{\ar@{-} (.1,-.7);(-.2,-.7)   *+!U{\Delta_0}}	
\end{xy}
\end{equation}
  \caption{The moduli space of the $\C^2/\Z_3$ model.}
  \label{fig:modsp}
\end{figure}

Now we have all the ingredients necessary for constructing the monodromy actions needed to generate the fractional branes. The toric fan for the moduli space of complexified Kahler forms was depicted in Fig.~\ref{fig:Z3}. The four maximal cones ${\mathcal C}_1$, ... , ${\mathcal C}_4$ correspond to the four distinguished phase points. The four edges correspond to curves in the moduli space. It is immediate to see these are all $\P^1$'s, at least topologically. For us the analytic structure will be important, and we need to be more careful here. The four curves connecting the different phase points are sketched in Fig.~\ref{fig:modsp}. The curves in question are weighted projective lines: ${\mathcal L}_1={\mathcal L}_2=\P^1(1,2)$,  ${\mathcal L}_3={\mathcal L}_4=\P^1(3,2)$. Since the singularities are in codimension one, these spaces in fact are not singular, and they are all isomorphic to $\P^1$.

In terms of the coordinates $(x_3,x_4)$ introduced in Eq.~(\ref{e:cpxc}) we have ${\mathcal L}_1\!: (x_4=0)$ and ${\mathcal L}_3\!: (x_3=0)$. The discriminant $\Delta_0$ intersects the four lines, and it is now clear from Eq.~(\ref{eq:dZ3}) that all intersections are transverse. We depicted this fact in Fig.~\ref{fig:modsp} using short segments.

When talking about monodromy there are two cases to be considered. One can loop around a divisor, i.e., real codimension two objects; or one can loop around a point inside a curve. Of course the two notions are not unrelated. Our interest will be the second type of monodromy: looping around a point inside a $\P^1$.

What we would like to write down is the monodromy inside ${\mathcal L}_3$ around the $\Z_3$ point. Since there is no direct approach to doing this, we follow an indirect way: both ${\mathcal L}_1$ and ${\mathcal L}_3$ are spheres, with three marked points, and we can compute the corresponding monodromies. Our approach is to go from the smooth point to the $\Z_3$ point by first ``moving'' inside  ${\mathcal L}_1$ and then ${\mathcal L}_3$. 

We start with ${\mathcal L}_1$, which has three distinguished points: the smooth point, ${\mathcal L}_1\cap \Delta_0$ and the $\Z_2$ point. Monodromy around the smooth point inside ${\mathcal L}_1$ is a large radius monodromy, and (\ref{e:ce1}) together with Prop.~\ref{p:lr} tell us that it is precisely $\ms{L}_{D_2}$. 

${\mathcal L}_1\cap \Delta_0$ is a  conifold-type point. Following \cite{Strominger:1995cz,Greene:1997tx} we know that it is the D-brane wrapping the shrinking cycle $C_3$ that should go massless at this point. But the mass depends on the central charge, which in turn is only a function of the K-theory class. At the K-theory level this fact has been verified \cite{DelaOssa:2001xk}. Conjecture~\ref{conj:a} then tells us the monodromy: $ \ms{T}_{i_*\O_{C_3}}$. 

Putting these two facts together we have the monodromy around the $\Z_2$ point inside ${\mathcal L}_1$:
\begin{equation}\label{e:m1} 
\ms{M}_{\Z_2}\, = \, \ms{T}_{i_*\O_{C_3}} \comp \, \ms{L}_{D_2}\,.
\end{equation}
We can immediately evaluate the kernel of this Fourier-Mukai functor
\begin{equation}\label{e:mk1}
\begin{split}
{\cal K}_{\Z_2}\, &= \, \Cone{(i_*\O_{C_3})^{\roof}\boxtimes i_*\O_{C_3} \longrightarrow \O_\Delta}\,\star \delta_*\O_X(D_2)\\
&= \, \Cone{(i_*\O_{C_3})^{\roof}\boxtimes i_*\O_{C_3}(1) \longrightarrow \delta_*\O_X(D_2) }.
\end{split}
\end{equation}
This expression will be useful when dealing with quantum symmetries.

Now we can  continue our march towards the $\Z_3$ point inside ${\mathcal L}_3$. Once again there are three distinguished points: the $\Z_3$ point, ${\mathcal L}_3\cap \Delta_0$ and the $\Z_2$ point. By the same token as before, monodromy around ${\mathcal L}_3\cap \Delta_0$ is $\ms{T}_{j_*\O_{C_4}}$. Of course we want the  monodromy around the $\Z_3$ point, so we need the monodromy around this $\Z_2$ point as well, which a priori has nothing to do with the previous $\Z_2$ monodromy inside ${\mathcal L}_1$. But the monodromy around the $\Z_2$ point is more subtle. Fig.~\ref{fig:modsp} in fact is quite misleading, since in reality the spheres ${\mathcal L}_1$ and ${\mathcal L}_3$ intersect transversely in 4-space. Moreover, the intersection point is an orbifold itself: $\C^2/\Z_2$. To see what happens, we need to work out the fundamental group of the complement of ${\mathcal L}_1$ and ${\mathcal L}_3$.\footnote{This situation is similar to the one analyzed in \cite{navigation}.}

Since the intersection point is a $\C^2/\Z_2$ orbifold, we surround it by the {\em lens space} $L=S^3/\Z_2$, instead of the usual sphere $S^3$.  ${\mathcal L}_1$ and ${\mathcal L}_3$ are both smooth curves and therefore intersect $L$ in unknotted circles. This way we reduced the problem of computing $\pi_1(\C^2/\Z_2-\{{\mathcal L}_1\cup {\mathcal L}_3\})$ to computing $\Pi=\pi_1(L-\{{\mathcal L}_1\cup {\mathcal L}_3\})$. To evaluate this consider the covering map $q\!:S^3\to L$, with free $\Z_2$ action, induced by the $\Z_2$ action on $\C^2$. The intersection of both ${\mathcal L}_1$ and ${\mathcal L}_3$ with $L$ lift under $q^{-1}$ to unknotted circles in $S^3$. These circles are linked once and thus $\pi_1(S^3-\{q^{-1}({\mathcal L}_1)\cup q^{-1}({\mathcal L}_3)\})=\Z\oplus\Z$.\footnote{An equivalent way of seeing this is to note that $\C^2-\{{\mathcal L}_1\cup {\mathcal L}_3\}$ is homotopic to $\C^*\times \C^*$, and that $S^1\times S^1$ is a deformation retract of $\C^*\times \C^*$. } 
The generators are the loops around $q^{-1}({\mathcal L}_1)$ and $q^{-1}({\mathcal L}_3)$, we call them $g_1$ and $g_2$.

Since $q$ is a normal cover we have a short exact sequence of abelian groups
\begin{equation} \label{eq:ext3}
\ses{\Z\oplus\Z}{\Pi}{\Z_2}\,.
\end{equation}
We can easily show that $\Pi=\Z\oplus\Z$ as well, by choosing a convenient fundamental domain, and two generators for $\Pi$: $l_1$ encircles ${\mathcal L}_1$, while $l_2$ goes from a basepoint to its antipodal. This second generator is a closed curve in  $L=S^3/\Z_2$ because of the quotienting, but it does not lift to $S^3$. Nevertheless, $2l_2$ {\em does} lift to $S^3$, and $q^{-1}(2l_2)=g_1+g_2$. In terms of the two basis $\left\langle g_1,g_2\right\rangle $ and $\langle l_1,l_2\rangle $ we have the non-trivial map in (\ref{eq:ext3}):
\begin{equation}
\xymatrix@1{\Z\oplus\Z 
\ar[rrr]^-{\left(\!\!
\begin{array}{rr}
1&-1\\
0&2
\end{array}
\right)} 
&&& \Z\oplus\Z}\,.
\end{equation}

Now we can continue our monodromy calculation. We claim that the loop around ${\mathcal L}_1\cap {\mathcal L}_3$ inside ${\mathcal L}_1$ is {\em homotopic} to the loop around ${\mathcal L}_1\cap {\mathcal L}_3$ inside ${\mathcal L}_3$. This statement is not to be taken literally though. Neither ${\mathcal L}_1$ nor ${\mathcal L}_3$ are part of the moduli space, so we are not looping inside them. What we have are loops that are infinitesimally close to such loops, but  lie outside ${\mathcal L}_1$ or ${\mathcal L}_3$. This distinction is usually irrelevant, but for us the singularity brings it to the forefront. What we need to do is to deform the loop inside ${\mathcal L}_1$ around ${\mathcal L}_1\cap {\mathcal L}_3$ so that it doesn't intersect ${\mathcal L}_1$ or ${\mathcal L}_3$, and similarly for the loop inside ${\mathcal L}_3$ around ${\mathcal L}_1\cap {\mathcal L}_3$. The reader can convince himself that the generic deformations are indeed both  homotopic to $l_2$. 

Therefore the monodromy inside ${\mathcal L}_3$ around the $\Z_3$ point is given by
\begin{equation}\label{e:m2}
\ms{M}_{\Z_3}\, = \, \ms{T}_{j_*\O_{C_4}}\comp \, \ms{M}_{\Z_2}
\, = \, \ms{T}_{j_*\O_{C_4}}\comp \,  \ms{T}_{i_*\O_{C_3}} \comp \, \ms{L}_{D_2}\,.
\end{equation}
The associated Fourier-Mukai  kernel is
\begin{equation}\label{e:mk2}
\begin{split}
{\cal K}_{\Z_3}\, &= \, \Cone{(j_*\O_{C_4})^{\roof}\boxtimes j_*\O_{C_4} \longrightarrow \O_\Delta}\,
\star {\cal K}_{\Z_2}\\
&= \, \operatorname{Cone}((j_*\O_{C_4})^{\roof}\boxtimes k_*\O_{C_3+C_4}(D_1+D_2) \longrightarrow \\
&\quad \longrightarrow \Cone{(i_*\O_{C_3})^{\roof}\boxtimes i_*\O_{C_3}(1) \rightarrow \delta_*\O_X(D_2)  })\,.
\end{split}
\end{equation}

\subsection{The $\Z_2$ kernel squared: $\ms{M}_{\Z_2}\comp  \ms{M}_{\Z_2}$}		\label{s:M22}

It is an interesting question to ask how the $\Z_2$ quantum symmetry of the partially resolved $\Z_2$ orbifold is realized in the derived category setup. Accordingly, we would like to compute the action of $(\ms{M}_{\Z_2})^2$ on a generic object. To get a feel for what to expect, we compute the Chern character of $(\ms{M}_{\Z_2})^2$ acting on the trivial bundle $\O_X$. 

Inspecting the form of $\ms{M}_{\Z_2}$ in Eq.~(\ref{e:m1}) we see that in order to compute $\ch((\ms{M}_{\Z_2})^2\O_X)$  some general properties might be of use. Taking the Chern character of both sides in Eq.~(\ref{e:refl}) one obtains \cite{ST:braid,navigation}:
\begin{equation}\label{e:monch}
\ch\big(\ms{T}_{\mathsf{A}}  (\mathsf{B})\big)
=\ch(\mathsf{B})-\langle\mathsf{A},\mathsf{B}\rangle\ch(\mathsf{A})\,,
\end{equation}
where $\langle\mathsf{A},\mathsf{B}\rangle$ is an Euler characteristic:
\begin{equation}
\langle\mathsf{A},\mathsf{B}\rangle =
\sum_i (-1)^i\dim\Ext_{\D(X)}^i(\mathsf{A},\mathsf{B})\,.
\end{equation}
The Grothendieck-Riemann-Roch theorem gives a useful way to compute this:
\begin{equation}\label{e:monch1}
\langle\mathsf{A},\mathsf{B}\rangle = \int_X \ch(\mathsf{A}^{\!\vee}) \ch(\mathsf{B})\td(X)\,.
\end{equation}

Using Eq.~(\ref{e:monch}) and Eq.~(\ref{e:monch1}) one obtains
\begin{equation}
\begin{split}
&\ch(\ms{M}_{\Z_2}\O_X) \quad = e^{D_2}\\
&\ch((\ms{M}_{\Z_2})^2\O_X)= e^{D_1}\\
\end{split}
\end{equation}
This suggests that $(\ms{M}_{\Z_2})^2$ acts like a large radius monodromy. A more involved computation shows that indeed on any object $x$
\begin{equation}\label{e:moni1}
\ch((\ms{M}_{\Z_2})^2\, x)= e^{D_1}\, \ch(x)\,.
\end{equation}
This result is a bit surprising, but a similar fact has been observed before \cite{navigation}. As discussed in Section~\ref{s:loops} the loop inside ${\mathcal L}_1$ and around ${\mathcal L}_1\cap {\mathcal L}_3$ after it is deformed becomes homotopic to $l_2$. $(\ms{M}_{\Z_2})^2$ thus corresponds to  $2l_2$, which lifts to the $S^3$, and encircles both ${\mathcal L}_1$ and ${\mathcal L}_3$. Eq.~(\ref{e:ce1}) tells us that monodromy around the divisor ${\mathcal L}_1$ is $\ms{L}_{\O_X(D_1)}$. Assuming that  monodromy around ${\mathcal L}_3$ is trivial, we have a perfect agreement with (\ref{e:moni1}). This is also consistent with the general statement that monodromy at an orbifold point has to be associated with $B$-field components other than the blow-up mode(s) of the orbifold, in this case $D_2$. From now on we assume that monodromy around ${\mathcal L}_3$ is trivial. We will see that this leads to consistent results.

Similarly we can ask about the $\Z_3$ quantum symmetry at the $\Z_3$ orbifold point. By the same argument, we could have a mismatch caused by monodromy around ${\mathcal L}_3$ and ${\mathcal L}_4$. But as argued in the previous paragraph, there is no monodromy around ${\mathcal L}_3$, and due to the symmetry of the problem there should be no monodromy around ${\mathcal L}_4$ either. We need to get $(\ms{M}_{\Z_3})^3=1$ on the nose. And indeed for a general object $x$, one has
\begin{equation}\label{e:moni2}
\ch((\ms{M}_{\Z_3})^3\, x)= \, \ch(x)\,.
\end{equation}
We note that $\ch((\ms{M}_{\Z_2})\, x)$ does  not have a simple expression, and similarly for $\ch((\ms{M}_{\Z_3})\, x)$ and $\ch((\ms{M}_{\Z_3})^2\, x)$.

Using the Fourier-Mukai technology we can go beyond the K-theory analysis and evaluate $(\ms{M}_{\Z_2})^2$ as it acts on the derived category. We can prove the following
\begin{prop}
In the notation of Prop.~\ref{p:lr} we have an equivalence of  functors:
\begin{equation}
(\ms{M}_{\Z_2})^2 \iso \ms{L}_{D_1}
\end{equation}
\end{prop}
\begin{proof}
We have seen in the previous section, Eq.~(\ref{e:mk1}), that the Fourier-Mukai kernel associated to monodromy around the $\Z_2$ point ${\cal C}_3$ is given by
\begin{equation}
{\cal K}_{\Z_2}\, = \, \Cone{(i_*\O_{C_3})^{\roof}\boxtimes i_*\O_{C_3}(1) \longrightarrow \delta_*\O_X(D_2) }\,.
\end{equation}
Let us make the following abbreviations:
\begin{equation}
A=(i_*\O_{C_3})^{\roof}\boxtimes i_*\O_{C_3}(1)\,, \quad B=\delta_*\O_X(D_2)\,, \quad
{\cal K}_{\Z_2}\, = \, \Cone{A \longrightarrow B }\,.
\end{equation}
According to Proposition~\ref{prop1} the kernel of $(\ms{M}_{\Z_2})^2$ is given by ${\cal K}_{\Z_2}\star {\cal K}_{\Z_2}$. As reviewed in Eq.~(\ref{trig:mor}), $-\star {\cal K}_{\Z_2}$ is an exact functor, and therefore
\begin{equation}
\begin{split}
{\cal K}_{\Z_2}\star {\cal K}_{\Z_2}\, &= \, \Cone{A \to B }\star {\cal K}_{\Z_2}	\\
&=\Cone{\Cone{A \star A\to A \star B } \longrightarrow \Cone{B \star A \to B \star B}}
\end{split}
\end{equation}

Using the  spectral sequences from Section~\ref{s:ss} one can show that
\begin{equation}
\begin{split}\label{eax1}
&A \star A = ((i_*\O_{C_3})^{\roof}\boxtimes i_*\O_{C_3}(1))^{\oplus 2} \qquad
A \star B = (i_*\O_{C_3})^{\roof}\boxtimes i_*\O_{C_3}(2) \\
&B \star A = (i_*\O_{C_3})^{\roof}(1)\boxtimes i_*\O_{C_3}(1) \qquad
B \star B = \delta_*\O_X(2D_2)\,.
\end{split}
\end{equation}

First we simplify 
\begin{equation}
\Cone{A \star A\to A \star B }=(i_*\O_{C_3})^{\roof}\boxtimes \Cone{i_*\O_{C_3}(1)^{\oplus 2}\to i_*\O_{C_3}(2)}\,.
\end{equation}
The Euler sequence for the tangent bundle of $\P^1$, or the Koszul  resolution of the complete intersection of two generic ``lines'' on $\P^1$, gives the short exact sequence
\begin{equation}\label{e:Euler}
\ses{\O_{\P^1}(-2)}{\O_{\P^1}(-1)^{\,\oplus 2}}{\O_{\P^1}}\,.
\end{equation}
This translates into the statement that
\begin{equation}
\Cone{ \O_{\P^1}(1)^{\,\oplus 2} \longrightarrow \O_{\P^1}(2) } = \O_{\P^1}[1]\,.
\end{equation}
Therefore
\begin{equation}
\Cone{A \star A\to A \star B }=(i_*\O_{C_3})^{\roof}\boxtimes i_*\O_{C_3} \,.
\end{equation}
Using this and the results in Eq.~(\ref{eax1}) we have\footnote{We underlined the $0$th position in the complex.}
\begin{equation}\label{ek2}
{\cal K}_{\Z_2}\star {\cal K}_{\Z_2}  = (i_*\O_{C_3})^{\roof}\boxtimes i_*\O_{C_3} \longrightarrow 
(i_*\O_{C_3}(-1))^{\roof}\boxtimes i_*\O_{C_3}(1) \longrightarrow \xymatrix{\poso{\delta_*\O_X(2D_2)}}\,.
\end{equation}
This is reminiscent of  {\em Beilinson's resolution of the diagonal} in $\P^d \times \P^d$ \cite{Bei:res}:
\begin{equation} \label{eq:beil}
0 \to \O_{\P^d}(-d) \boxtimes \Omega_{\P^d}^d (d) \to \ldots \to
\O_{\P^d}(-1) \boxtimes \Omega_{\P^d}^1 (1) \to \O_{\P^d \times \P^d} \to \O_{\P^d} \to 0,
\end{equation}
where $\Omega_{\P^d}^i$ is the sheaf of holomorphic $i$-forms on $\P^d.$ Beilinson's resolution for $C_3=\P^1$ is in fact very simple:
\begin{equation} \label{eq:bei2}
\ses{\O_{C_3}(-1) \boxtimes_{C_3} \O_{C_3}(-1)}{\O_{C_3} \boxtimes_{C_3} \O_{C_3}}{{j_3}_*\O_{C_3} }
\end{equation}
where $j_3\!: {C_3} \hookrightarrow {C_3}\!\times\! {C_3}$ is the diagonal map. The notation $\boxtimes_{C_3}$ makes explicit where the exterior product is considered.

But this is a short exact sequence on $C_3  \!\times\!C_3$, while Eq.~(\ref{ek2}) is a statement in $X\!\times\!X$. Fortunately we can relate the two:
\begin{lemma}
Let $Z$ be a subvariety of the variety $X$, and let $i\! : Z \hookrightarrow X$ be the embedding. For two sheaves $\c{A}$ and $\c{B}$, on $Z$ one has that 
\begin{equation}\label{en6}
i_*\c{A} \boxtimes_X i_*\c{B} = (i\!\times\! i)_*\, ( \c{A} \boxtimes_Z \c{B})\,.
\end{equation}
\end{lemma}
\begin{proof}[Proof of the Lemma]
First we specify some notation: $p_i$ denotes projection on the $i$th factor of $X\!\times\!X$. Similarly $s_i$ projects on the $i$th factor of $Z\!\times\! Z$.

By definition then $i_*\c{A} \boxtimes_X i_*\c{B} = p_2^*i_*\c{A} \otimes p_1^*i_*\c{B}$. But $p_j^*\comp i_* = (i\!\times\!i)_* \comp s_j^*$, as follows from the following fiber square:
\begin{equation}
\xymatrix@C=15mm{ Z\!\times\! Z \ar[r]^-{s_i} \ar[d]_{i\!\times\!i} & Z \ar[d]^i \\
 		  X\!\times\!X \ar[r]^-{p_i} 				   	& 	X 		}
\end{equation}
Thus $p_2^*i_*\c{A} \otimes p_1^*i_*\c{B} = (i\!\times\!i)_* s_2^* \c{A} \otimes p_1^*i_*\c{B}$. By the projection formula this is $(i\!\times\!i)_* (s_2^*\c{A}  \otimes (i\!\times\!i)^* p_1^*i_*\c{B})$. Now $(i\!\times\!i)^* p_1^*= (p_1\comp (i\!\times\!i) )^*$. Next we can use the fact that the above diagram is commutative and rewrite further: $(p_1\comp (i\!\times\!i) )^*=(i\comp s_1)^*=s_1^* i^*$. Since $\c{A} \boxtimes_Z \c{B} = s_2^*\c{A}  \otimes s_1^*\c{B}$ this completes the proof of the lemma.
\end{proof}


Returning to Eq.~(\ref{ek2}), let us look at the first two terms in the complex:
\begin{equation}\label{ek22}
 (i_*\O_{C_3})^{\roof}\boxtimes i_*\O_{C_3} \longrightarrow 
(i_*\O_{C_3}(-1))^{\roof}\boxtimes i_*\O_{C_3}(1) \,.
\end{equation}
The dual of the short exact sequence
\begin{equation}	\label{ek221}
\xymatrix{0 \ar[r] & \O_X(-D) \ar[r] & \O_X \ar[r] & \O_{D} \ar[r] & 0}
\end{equation}
shows that 
\begin{equation}
\O_{D}^{\,\roof}=\O_{D}(D)[-1]\,.
\end{equation}
Therefore (\ref{ek22}) can be rewritten as
\begin{equation}
\{ i_*\O_{C_3}(-2)\boxtimes i_*\O_{C_3} \longrightarrow 
i_*\O_{C_3}(-1)\boxtimes i_*\O_{C_3}(1)\}[-1] \,.
\end{equation}
Using (\ref{en6}) this becomes
\begin{equation}
(i\!\times\!i)_* \{ \O_{C_3}(-2)\boxtimes \O_{C_3} \longrightarrow 
\O_{C_3}(-1)\boxtimes \O_{C_3}(1)\}[-1] \,,
\end{equation}
where now the $\boxtimes$ is on $C_3$ rather than $X$. To bring the last expression to the Beilinson form we rewrite it as
\begin{equation}\label{ek23}
(i\!\times\!i)_* \{ [ \O_{C_3}(-1)\boxtimes \O_{C_3}(-1) \longrightarrow 
\O_{C_3}\boxtimes \O_{C_3}]\otimes s_2^*\O_{C_3}(-1)\otimes s_1^*\O_{C_3}(1)\}[-1] \,,
\end{equation}
where $s_i\!: {C_3}\!\times\! {C_3}\to {C_3}$ is projection on the $i$th factor. Now using the Beilinson resolution from Eq.~(\ref{eq:bei2}) this becomes
\begin{equation}
(i\!\times\!i)_* \{ {j_3}_*\O_{C_3} \otimes s_2^*\O_{C_3}(-1)\otimes s_1^*\O_{C_3}(1)\}[-1] \,.
\end{equation}
Using the projection formula we have:
\begin{equation}\label{ek213}
(i\!\times\!i)_* {j_3}_*\{ \O_{C_3} \otimes {j_3}^*s_2^*\O_{C_3}(-1)\otimes {j_3}^*s_1^*\O_{C_3}(1)\}[-1] \,.
\end{equation}
By construction $s_i\comp j_3 = \id$. From the commutative diagram
\begin{equation}
\xymatrix@C=15mm{
*++{C_3} \ar@{^{(}->}[r]^-{j_3} \ar@{^{(}->}[dr]_{i} & *++{C_3  \!\times\!C_3} \ar@{^{(}->}[r]^{i\!\times\!i} &  *++{X\!\times\!X} \\
		   				   	& 	*++{X} \ar@{^{(}->}[ur]_{\delta}	&	
}
\end{equation}
we also have that $(i\!\times\!i) \comp j_3 = \delta \comp i$. This simplifies  (\ref{ek213}) and it becomes simply $\delta_*i_*\O_{C_3}[-1]$. Using this (\ref{ek2}) is nothing but 
\begin{equation}
{\cal K}_{\Z_2}\star {\cal K}_{\Z_2}  = 
\Cone{\delta_*i_*\O_{C_3}[-1] \longrightarrow \delta_*\O_X(2D_2)}=\delta_* \Cone{i_*\O_{C_3}[-1] \longrightarrow \O_X(2D_2)}\,.
\end{equation}
The Chern character of this is $e^{D_1}$, and a bit more work shows that indeed 
\begin{equation}
{\cal K}_{\Z_2}\star {\cal K}_{\Z_2}  =\delta_* \O_X(D_1)\,.
\end{equation}
To see this note that $\Cone{i_*\O_{C_3}[-1] \longrightarrow \O_X(2D_2)}=\Cone{i_*\O_{C_3}[-1] \longrightarrow \O_X(2D_2-D_1)}\otimes \O_X(D_1)$. The relations (\ref{se1}) imply that $2D_2-D_1=-D_3$. Using the exact triangle storming from (\ref{ek221})  shows that $\Cone{i_*\O_{C_3}[-1] \longrightarrow \O_X(-D_3)}= \O_X$.

Finally, the use of Prop. \ref{p:lr} then completes the proof of the theorem.
\end{proof}

Based on (\ref{e:moni2}) and our experience with $\ms{M}_{\Z_2}$, we conjecture that ${\cal K}_{\Z_3}\star {\cal K}_{\Z_3}\star{\cal K}_{\Z_3}  =\delta_* \O_X$, but judging by the above  proof and the fact that ${\cal K}_{\Z_3}$ is much more complicated than ${\cal K}_{\Z_2}$, we didn't even attempt proving this. Nevertheless, we will have evidence for the conjecture in Section~\ref{s:3rdtime}

\section{The $\C^2/\Z_3$ fractional branes}	\label{s:cnfract}

In this section we use the $\Z_3$ monodromy action found in the previous section to generate a collection of fractional branes, and study some of their properties. As a starting point we need to know one of the fractional branes. We assume that the D5-brane wrapping the exceptional divisor $C_3$ is one of the fractional branes. This is a natural  assumption as long as we do not make any claims about the rest of the fractional branes. It is reasonable to expect that by various monodromy transformations any one of the fractional branes can be brought to this form. Instead of guessing the the other two fractional branes, we look at the orbit of this D5-brane under the $\Z_3$ monodromy action. In the quiver language the fractional branes are the simple representations of the quiver, and are mapped into each other under the $\Z_3$ quantum symmetry. Therefore, the  fractional branes will necessarily form a length three orbit of the $\Z_3$ monodromy, which is an incarnation of the $\Z_3$ quantum symmetry.

By the same token we could have chosen $C_4$ to be the fractional branes to start with. The $\C^2/\Z_3$ geometry is completely symmetric with respect to   $C_3$ and  $C_4$, and therefore we expect that whichever we start with, the other one will show up in the orbit. This is precisely what we are going to find.

\subsection{Generating fractional branes}

We start by recalling Eq.~(\ref{e:m2}), which gives us the form of the $\Z_3$ monodromy $\ms{M}_{\Z_3}$:
\begin{equation}\label{e:m3}
\ms{M}_{\Z_3}\, = \, \ms{T}_{j_*\O_{C_4}}\comp \,  \ms{T}_{i_*\O_{C_3}} \comp \, \ms{L}_{D_2}
\end{equation}
By the assumption made above the $1$st fractional brane is $i_*\O_{C_3}$, and the other two are  $\ms{M}_{\Z_3}(i_*\O_{C_3})$ and $(\ms{M}_{\Z_3})^2 (i_*\O_{C_3})$. We start out by computing $\ms{M}_{\Z_3}(i_*\O_{C_3})$.

\subsubsection{Computing $\ms{M}_{\Z_3}(i_*\O_{C_3})$}

The first step is quite trivial:
\begin{equation}
\xymatrix{i_*\O_{C_3} \ar[rr]^{\ms{L}_{D_2}} && i_*\O_{C_3}(1)}\,.
\end{equation}
We can act on this with $\ms{T}_{\O_{C_3}}$, and use the fact that $\R i_*$ is a triangulated functor, to obtain
\begin{equation}
\xymatrix{i_*\O_{C_3}(1) \ar[rr]^-{\ms{T}_{\O_{C_3}}} && 
\Cone {i_*\O_{C_3}^{\,\oplus 2} \longrightarrow i_*\O_{C_3}(1) }} =
i_* \Cone {\O_{C_3}^{\,\oplus 2} \longrightarrow \O_{C_3}(1) }\,.
\end{equation}
The intermediate steps above involved using the spectral sequence (\ref{SS1}), but we suppress the details. Now we can use the Euler sequence (\ref{e:Euler}) and simplify
\begin{equation}
\xymatrix{i_*\O_{C_3}(1) \ar[rr]^-{\ms{T}_{\O_{C_3}}} && i_*\O_{C_3}(-1)[1]}\,.
\end{equation}
The final leg involves using the spectral sequence (\ref{SS3}) and results in
\begin{equation}
\xymatrix{i_*\O_{C_3}(-1)[1] \ar[rr]^-{\ms{T}_{\O_{C_4}}} && 
\Cone {j_*\O_{C_4} \longrightarrow i_*\O_{C_3}(-1)[1] }} = k_*\O_{C_3+C_4}[1]\,.
\end{equation}
Here in the last equality we used the exact triangle 
\begin{equation}\label{e:bpvdw}
\xymatrix{i_*\O_{C_3}(-1) \ar[r] & k_*\O_{C_3+C_4} \ar[r] &  j_*\O_{C_4} \ar[r] & i_*\O_{C_3}(-1)[1] }
\end{equation}
stemming from the short exact sequence:\footnote{The intuition behind this formula is easy to understand. The inclusion map of $D$ into $C+D$ allows us to restrict functions from $C+D$ to $D$. This is the map $\O_{C+D} \rightarrow \O_{D}$. The kernel of this map consists of those functions on $C$ that vanish at the intersection point with $D$: $\O_{C}(-D)$. For a rigorous proof see \cite{barthpvdv}.}
\begin{equation}\label{e:bpvdw1}
\xymatrix{0 \ar[r] & \O_{C}(-D) \ar[r] & \O_{C+D} \ar[r] & \O_{D} \ar[r] & 0}\,.
\end{equation}
Thus the $2$nd fractional brane is the D5-brane that wraps both exceptional divisors, $C_3$ and $C_4$.

\subsubsection{Computing $(\ms{M}_{\Z_3})^2(i_*\O_{C_3})$}

To determine the $3$rd fractional brane we apply the $\Z_3$ monodromy again. Now the starting point is the second fractional brane  $k_*\O_{C_3+C_4}(D_2)[1]$ from the previous section:
\begin{equation}
\xymatrix{k_*\O_{C_3+C_4}[1] \ar[rr]^-{\ms{L}_{D_2}} &&  k_*\O_{C_3+C_4}(D_2)[1]\,.}
\end{equation}

For the next step we need $\RHom_X(i_*\O_{C_3}, k_*\O_{C_3+C_4}(D_2)[1])$. We will determine these $\Ext$ groups in two different ways. The first method will use the cohomology long exact sequence associated to an exact triangle. The second method will use the spectral sequence derived in Appendix~\ref{a:1}. Although the first method is a priori more straightforward, we will see that once the general spectral sequence result is established, it  is much more efficient. The fact that the two methods give the same result provides a consistency check for our calculations.

We start with the  long exact sequence associated to the exact triangle (\ref{e:bpvdw}) once the covariant functor $\Ext^i_X(i_*\O_{C_3},-)$ is applied to it:
\begin{equation}\label{e:lesd}
\Ext^i_X(i_*\O_{C_3}, j_*\O_{C_4} )\longrightarrow  \Ext^{i+1}_X(i_*\O_{C_3}, i_*\O_{C_3})\longrightarrow  
\Ext^i_X(i_*\O_{C_3}, k_*\O_{C_3+C_4}(D_2)[1])
\end{equation}
The spectral sequence (\ref{SS3}) tells us that $\Ext^i_X(i_*\O_{C_3}, j_*\O_{C_4} )=\delta_{i,1}$, while using the spectral sequence  (\ref{SS1}) gives us \footnote{The same result for $\Ext^i_X(i_*\O_{C_3}, i_*\O_{C_3}) $  follows if we use Prop.~3.15 of \cite{ST:braid}, which guarantees that $i_*\O_{C_3}$ is a spherical object in the sense of Definition~\ref{def:spherical}. The same is true for $j_*\O_{C_4}$.}
\begin{equation}\label{e:lesd1}
\Ext^i_X(i_*\O_{C_3}, i_*\O_{C_3}) = \left\{ 
\begin{array}{rr}
\C & \mbox{for $i=0,2$}	\\
0   & \mbox{otherwise}.
\end{array}	\right.
\end{equation}
Using these two facts, the long exact sequence  (\ref{e:lesd}) tells us that $\Ext^i_X(i_*\O_{C_3}, k_*\O_{C_3+C_4}(D_2)[1])=\delta_{i,-1}$.

The same result can be obtained much quicker, if we apply the spectral sequence (\ref{eq:dcx}) to our case. What we need to compute is $$\Ext^i_X(i_*\O_{C_3}, k_*\O_{C_3+C_4}(D_2)[1])=\Ext^{i+1}_X(i_*\O_{C_3}(-1), k_*\O_{C_3+C_4}).$$ 
By Serre duality this becomes $\Ext^{1-i}_X( k_*\O_{C_3+C_4}, i_*\O_{C_3}(-1))$. The spectral sequence (\ref{eq:dcx}) then reads
\begin{equation}
\begin{xy}
\xymatrix@C=2mm{
   & \H^p (\P^1,  \O(-2)) & \\
   & \H^p (\P^1,  \O(-1)) & }
\save="x"!LD+<-3mm,0pt>;"x"!RD+<0pt,0pt>**\dir{-}?>*\dir{>}\restore
\save="x"!LD+<0pt,-3mm>;"x"!LU+<0pt,-2mm>**\dir{-}?>*\dir{>}\restore
\save!CD+<0mm,-4mm>*{p}\restore
\save!CL+<-3mm,0mm>*{q}\restore
\end{xy}
\begin{xy}
\xymatrix@C=10mm{
    &  &\\
    &= &\\
     &&\\}
\end{xy}
\begin{xy}
\xymatrix@C=15mm{
    0 & \C \\
    0 & 0 }
\save="x"!LD+<-3mm,0pt>;"x"!RD+<0pt,0pt>**\dir{-}?>*\dir{>}\restore
\save="x"!LD+<0pt,-3mm>;"x"!LU+<0pt,-2mm>**\dir{-}?>*\dir{>}\restore
\save!CD+<0mm,-4mm>*{p}\restore
\save!CL+<-3mm,0mm>*{q}\restore
\end{xy}
\end{equation}
and therefore $\RHom_X(i_*\O_{C_3}, k_*\O_{C_3+C_4}(D_2)[1])=\C[1]$, as we saw before.

With the ``algebra'' out of the way, we can head back to monodromy and establish that 
\begin{equation}
\xymatrix{ k_*\O_{C_3+C_4}(D_2)[1] \ar[rr]^-{\ms{T}_{\O_{C_3}}} && 
\Cone{i_*\O_{C_3}[1]\to k_*\O_{C_3+C_4}(D_2)[1]}\,.}
\end{equation}

The RHS can be simplified using (\ref{e:bpvdw}), and gives a simple answer: 
\begin{equation}
\Cone{i_*\O_{C_3}[1]\to k_*\O_{C_3+C_4}(D_2)[1]} =  j_*\O_{C_4}[1]\,.
\end{equation}
Therefore the last step of the computation involves $\ms{T}_{\O_{C_4}}(j_*\O_{C_4})[1]$. Here we can use a more general result
\begin{lemma}\label{lemma:41}
If $\mathsf{A}$ is an $n$-spherical object, then $\ms{T}_{\mathsf{A}} (\mathsf{A}) = \mathsf{A}[1-n]$.
\end{lemma}
\begin{proof}
By the definition of $\ms{T}_{\mathsf{A}} (\mathsf{A})$ and the $n$-sphericity of $\mathsf{A}$ one has
\begin{equation}\label{eq:Ocon}
\begin{split}
 & \Cone{\RHom_{\D(X)}(\mathsf{A},\mathsf{A})\Ltensor\mathsf{A}\to\mathsf{A}}
  = \Cone{(\xymatrix{\poso{\C}}\to 0\to \ldots \to 0 \to \C)\Ltensor\mathsf{A}\to\mathsf{A}} \\
  &= \Cone{\mathsf{A}\oplus \mathsf{A}[-n] \longrightarrow \mathsf{A}}
  \cong \mathsf{A}[1-n]\,,
\end{split}
\end{equation}
\end{proof}

For $n=2$ and $\mathsf{A}=j_*\O_{C_4}$\footnote{Eq.~(\ref{e:lesd1}) guarantees that $j_*\O_{C_4}$ is 2-spherical.}
the lemma gives $\ms{T}_{j_*\O_{C_4}}(j_*\O_{C_4})= j_*\O_{C_4}[-1]$, and thus 
\begin{equation}
\xymatrix{j_*\O_{C_4}[1]  \ar[rr]^-{\ms{T}_{\O_{C_4}}} &&  j_*\O_{C_4}\,.}
\end{equation}
This establishes $j_*\O_{C_4}$ as the $3$rd fractional brane.

\subsection{Consistency check}	\label{s:3rdtime}

In order to test the $\Z_3 $ quantum symmetry conjecture we verify the closure of the $\Z_3 $ orbit $i_*\O_{C_3}$, $k_*\O_{C_3+C_4}(D_2)[1]$, $j_*\O_{C_4}$  under the $\Z_3 $ monodromy. The first two steps are:
\begin{equation}
\xymatrix{j_*\O_{C_4} \ar[rr]^-{\ms{L}_{D_2}} && j_*\O_{C_4}  \ar[rr]^-{\ms{T}_{\O_{C_3}}} && 
\Cone{i_*\O_{C_3}[-1]\longrightarrow j_*\O_{C_4}}\,.}
\end{equation}
The next step is to compute the action of $\ms{T}_{\O_{C_4}}$ on this.
Since $\ms{T}_{\O_{C_4}}$ is an exact functor, i.e. a functor between triangulated categories, we have that
\begin{equation}
\ms{T}_{\O_{C_4}}\Cone{i_*\O_{C_3}[-1]\longrightarrow j_*\O_{C_4}}=
\Cone{ \ms{T}_{\O_{C_4}} (i_*\O_{C_3})[-1]\longrightarrow \ms{T}_{\O_{C_4}} (j_*\O_{C_4})}\,.
\end{equation}
But
\begin{equation}
 \ms{T}_{\O_{C_4}} (i_*\O_{C_3})=\Cone{j_*\O_{C_4}[-1] \longrightarrow i_*\O_{C_3}}=k_*\O_{C_3+C_4}(D_2)\,,
\end{equation}
and by Lemma~\ref{lemma:41}
\begin{equation}
\ms{T}_{\O_{C_4}} (j_*\O_{C_4})=j_*\O_{C_4}[-1]\,.
\end{equation}
Therefore
\begin{equation}
\ms{T}_{\O_{C_4}}\Cone{i_*\O_{C_3}[-1]\longrightarrow j_*\O_{C_4}}=
\Cone{k_*\O_{C_3+C_4}(D_2)\longrightarrow j_*\O_{C_4}}[-1]=i_*\O_{C_3}\,. 
\end{equation}
In other words $(\ms{M}_{\Z_3})^3(i_*\O_{C_3})=i_*\O_{C_3}$. A similar computation shows that $(\ms{M}_{\Z_3})^3(j_*\O_{C_4})=j_*\O_{C_4}$. These two facts provide evidence for the $(\ms{M}_{\Z_3})^3 = \id$ conjecture.

\section{Generalizations}

\subsection{Connection with the McKay correspondence}

The original version of the McKay correspondence \cite{McKay} relates the representations of a finite subgroup  $\Gamma$  of $\SL(2,\C)$ to the cohomology of the minimal resolution of the Kleinian singularity $\C^2/\Gamma$. Gonzalez-Sprinberg and Verdier \cite{SprinbergVerdier} reinterpreted the McKay correspondence as a K-theory isomorphism, observing that the representation ring of $\Gamma$ is the same as the $\Gamma$-equivariant K-theory of $\C^2$. But even at this point a deeper understanding of the correspondence was lacking, all the results were based on a case by case analysis. 

A solid understanding of the McKay correspondence culminated with the work of Bridgeland, King and Reid \cite{Mukai:McKay}, who showed that in dimensions two and three the McKay correspondence is an equivalence of two very different derived categories.\footnote{The Bridgeland-King-Reid proof formally generalizes to higher dimensions, but then one needs more information about the existence of crepant resolutions, and in particular about Nakamura's $\Gamma$-Hilbert scheme.} 
In fact the seeds of the Bridgeland-King-Reid construction were already implicit in the work of Gonzalez-Sprinberg and Verdier \cite{SprinbergVerdier}, which inspired Kapranov and Vasserot \cite{Kapranov:Vasserot} to prove the derived McKay correspondence for $\C^2/\Gamma$  prior to the general Bridgeland-King-Reid proof. 

As we will see, Kapranov and Vasserot implicitly provide a collection of fractional branes for arbitrary $\Gamma \subset \SL(2,\C)$, which is different from what we obtained by monodromy for $\Z_3$. Let us review their construction.

First of all, we have the covering map $p\!: \C^2\longrightarrow \C^2/\Gamma$. Then we have the map $\tilde{p}\!: X\longrightarrow \C^2/\Gamma$ corresponding to the resolution of singularities. Using these two maps we can consider the fiber product $Y$ of $\C^2$ and $X$ over $\C^2/\Gamma$:
\begin{equation}
\xymatrix@C=15mm{
X\!\times\!\C^2 \ar@/^5mm/[rrd]^{p_2}\ar@/_5mm/[ddr]^{p_1}  & &   \\
 & Y \ar[r]^{q_2} \ar[d]_{q_1} & \C^2 \ar[d]^{p}  \\
 & X \ar[r] ^{\tilde{p}} & \C^2/\Gamma	}
\end{equation}
On the same diagram we depicted the projection maps $p_1$ and $p_2$ of the product $X\!\times\!\C^2$. $Y$ is in fact an incidence subscheme in $X\!\times\!\C^2$, and $q_i$ is the restriction of $p_i$ to $Y$.

Let $\Coh^\Gamma(\C^2)$ be the category of $\Gamma$-equivariant coherent sheaves on $\C^2$, and $\Coh(X)$ be the category of coherent sheaves on $X$. Kapranov and Vasserot define two functors: 
\begin{equation}\label{eq:psiphi}
\begin{split}
&\Phi\!: \D(\Coh^\Gamma(\C^2))\longrightarrow \D(\Coh(X))	\qquad \Phi(\c F) = (\R q_{1*} \L q _{2}^* \c F)^\Gamma \\
&\Psi\!: \D(\Coh(X)) \longrightarrow  \D(\Coh^\Gamma(\C^2))\qquad \Psi(\c G) = \R p_{2*} \R\!\sHom(\O_Y , p_1^*\c  G).
\end{split}
\end{equation}
The main result of their paper shows that $\Phi$ and $\Psi$  are mutually inverse equivalences of categories. Moreover, they also determine the images under  $\Phi$  of some special objects in $\D(\Coh^\Gamma(\C^2))$. To define what these objects are, let us recall that a finite-dimensional representation $V$ of $\Gamma$ gives rise to two equivariant sheaves on $\C^2$: 
\begin{enumerate}
\item the skyscraper sheaf $V^!$, whose fiber at 0 is V and all the other fibers vanish, and 
\item the locally free sheaf $\tilde{V} = V \otimes_{\O_{\C}} \O_{\C^2}$. 
\end{enumerate}

There is a one-to-one correspondence between the representations of the McKay quiver and  the category of $\Gamma$-equivariant coherent sheaves on $\C^2$ (for a review of this see, e.g., \cite{Paul:TASI2003}). In the language of quiver representations the fractional branes are the simple objects, i.e. with no sub-objects; the representations with all but one node assigned the trivial vector space, and all arrows are assigned the $0$ morphisms. The non-trivial node is assigned the vector space $\C$. Under this equivalence, the simple objects correspond to the skyscraper sheaves $\pi^!$, which are assigned to the irreducible representations $\pi$ of $\Gamma$. Therefore the fractional branes are the $\Gamma$-equivariant sheaves $\pi^!$.

Let $C_i$ be the exceptional divisors of the crepant resolution of $\C^2/\Gamma$, for $\Gamma$ a subgroup of $\SL(2,\C)$. A theorem in Section~2.3 of \cite{Kapranov:Vasserot} asserts that the image of the collection $\{\pi^!\! : \mbox{$\pi $ irrep of $\Gamma$}\}$ under the functor $\Phi$ from (\ref{eq:psiphi}) is
\begin{equation}\label{e:colMcKayN}
\O_{\sum_j C_j} \quad \mbox{and} \quad \O_{C_i}(-1)[1]\,.
\end{equation}
Specializing this result to the $\C^2/\Z_3$ case, and relabeling, gives the collection 
\begin{equation}\label{e:colMcKay}
\O_{C_3+C_4}\,, \quad \O_{C_3}(-1)[1]\,, \quad \O_{C_4}(-1)[1]\,.
\end{equation}
This collection is to be contrasted with the one obtained in Section~\ref{s:cnfract}:
\begin{equation}\label{e:colEn}
\O_{C_3+C_4}[1]\,, \quad \O_{C_3}\,, \quad \O_{C_4}\,.
\end{equation}

Next we show that both collections give what is expected of them, and also elucidate their connection. We will also investigate to what extent do these collections generalize for $\C^2/\Z_n$ with $n>3$.

\subsubsection{The quiver}

Let us start with the collection  (\ref{e:colEn}) obtained by monodromy. Using the spectral sequence (\ref{eq:dcx}) it is immediate that $\Ext^i_X( k_*\O_{C_3+C_4}[1],i_*\O_{C_3})=\delta_{i,1}$. By Serre duality this also means that $\Ext_X^i( i_*\O_{C_3}, k_*\O_{C_3+C_4}[1])=\delta_{i,1}$. A similar result holds for $\O_{C_4}$ as well. In fact all the results are invariant under the interchange of $C_3$ and $C_4$. Finally, the  spectral sequence (\ref{SS3}) shows that $\Ext_X^i(j_*\O_{C_4},i_*\O_{C_3})=\delta_{i,1}$. 

\begin{figure}[h]
\begin{equation}\nonumber
\begin{xy} <1.5mm,0mm>:
  (0,0)*{\bullet}="a" ,(10,17)*{\bullet}="c" ,(20,0)*{\bullet}="b"
  ,(0,-4)*{\scriptstyle \O_{C_3}},(17,17)*{\scriptstyle \O_{C_3+C_4}[1]},(20,-4)*{\scriptstyle \O_{C_4}}
  \ar@{-}@/^1mm/|*\dir{<}"a";"b"
  \ar@{-}@/_1mm/|*\dir{>}"a";"b"   
  \ar@{-}@/^1mm/|*\dir{<}"a";"c"
  \ar@{-}@/_1mm/|*\dir{>}"a";"c"
  \ar@{-}@/^1mm/|*\dir{<}"c";"b"
  \ar@{-}@/_1mm/|*\dir{>}"c";"b"
  \ar@{->}@`{(-3,-3),(-5,-1),(-5,1),(-3,3)} "a";"a"
  \ar@{->}@`{(23,-3),(25,-1),(25,1),(23,3)} "b";"b"
  \ar@{->}@`{(13,20),(11,22),(9,22),(7,20)} "c";"c"
\end{xy}
\qquad\qquad
\begin{xy} <1.5mm,0mm>:
  (0,0)*{\bullet}="a" ,(10,17)*{\bullet}="c" ,(20,0)*{\bullet}="b"
  ,(0,-4)*{\scriptstyle \O_{C_3}(-1)[1]},(17,17)*{\scriptstyle \O_{C_3+C_4}},(20,-4)*{\scriptstyle \O_{C_4}(-1)[1]}
  \ar@{-}@/^1mm/|*\dir{<}"a";"b"
  \ar@{-}@/_1mm/|*\dir{>}"a";"b"   
  \ar@{-}@/^1mm/|*\dir{<}"a";"c"
  \ar@{-}@/_1mm/|*\dir{>}"a";"c"
  \ar@{-}@/^1mm/|*\dir{<}"c";"b"
  \ar@{-}@/_1mm/|*\dir{>}"c";"b"
  \ar@{->}@`{(-3,-3),(-5,-1),(-5,1),(-3,3)} "a";"a"
  \ar@{->}@`{(23,-3),(25,-1),(25,1),(23,3)} "b";"b"
  \ar@{->}@`{(13,20),(11,22),(9,22),(7,20)} "c";"c"
\end{xy}
\end{equation}
  \caption{The  quivers for the two fractional $\C^2/\Z_3$ collections.}
  \label{f:c2z3}
\end{figure}
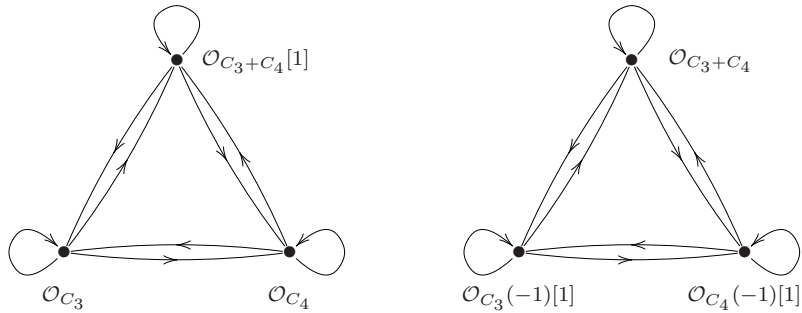

These results give us the well-known $\C^2/\Z_3$  quiver, depicted in Fig.~\ref{f:c2z3}.
Note that in writing down the quiver we used the result of Eq.~(\ref{e:lesd1}) to draw the loops corresponding to the adjoint fields. This took care only of $\O_{C_3}$ and $\O_{C_4}$. For $\Ext_X^i( k_*\O_{C_3+C_4}[1], k_*\O_{C_3+C_4}[1])$ we can use the exact sequence (\ref{e:bpvdw})  and apply the covariant functor $\Hom_{\D(X)}( k_*\O_{C_3+C_4}[1], -)$ to it, and get the same answer:
\begin{equation}\label{e:bpvdw11}
\Ext_X^i( k_*\O_{C_3+C_4}, k_*\O_{C_3+C_4})=\left\{
\begin{array}{rr}
\C & \mbox{for $i=0,2$}	\\
0	& \mbox{for $i\neq 0,2$}	 
\end{array}	\right.
\end{equation}

A similar computation shows that  the collection  (\ref{e:colMcKay}) gives the same quiver. Therefore we have two potential sets of fractional branes. The next thing we need to check is whether their central charges add up to that of the D3-brane. The various Chern characters are easiest to compute from (\ref{e:bpvdw1}). In general, for  $\C^2/Z_n$, we have:\footnote{For brevity every embedding map is denoted by $i$.}
\begin{equation}
\ch(i_*\O_{\sum C_j}) =p + \sum_{j=1}^{n-1}C_j\,, \quad
\ch(i_*\O_{C_i}(-1)) =  C_i\,, \quad
\ch(i_*\O_{C_i}) =C_i+p\,,
\end{equation}
where $p$ denotes the class of a point. Therefore
\begin{equation}
\ch(i_*\O_{\sum C_j}) + \sum_{i=1}^{n-1}\ch(i_*\O_{C_i}(-1)[1]) = \ch(\O_p)\,.
\end{equation}
On the other hand
\begin{equation}\label{e:wr4}
\ch(i_*\O_{\sum C_j}[1]) + \sum_{i=1}^{n-1}\ch(i_*\O_{C_i}) = (n-2)\ch(\O_p)\,.
\end{equation}
This shows that both collections (\ref{e:colMcKay}) and (\ref{e:colEn}) give fractional branes for $n=3$. 

Naturally, one would still need to check that the central charge, and hence the mass of these branes, is a third of the D0-brane  central charge {\em at the} $\Z_3$ point in moduli space. This can be done very easily using local mirror symmetry and the expression for the central charges in terms of the periods given in \cite{Skarke:2001ki}. The reason why this works is that the  central charges are determined by the large volume asymptotics, which depend only on the Chern character of the brane.

A simple computation shows that the collection (\ref{e:colMcKayN}) gives the $\C^2/Z_n$ quiver for any $n\geq 2$, not only for $n=3$:
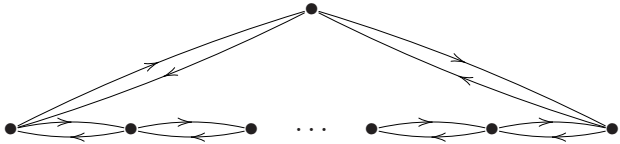
\begin{equation}\nonumber
\begin{xy} <0.8mm,0mm>:
  (0,0)*{\bullet}="a",(50,20)*{\bullet}="c",(100,0)*{\bullet}="b",(40,0)*{\bullet}="d"
  ,(60,0)*{\bullet}="e",(20,0)*{\bullet}="f",(80,0)*{\bullet}="g"
  ,(50,0)*{\ldots}
  \ar@{-}@/^1mm/|*\dir{>} "a";"f"
  \ar@{-}@/_1mm/|*\dir{<} "a";"f"
  \ar@{-}@/^1mm/|*\dir{>} "a";"c"
  \ar@{-}@/_1mm/|*\dir{<} "a";"c"
  \ar@{-}@/^1mm/|*\dir{>} "c";"b"
  \ar@{-}@/_1mm/|*\dir{<} "c";"b"
  \ar@{-}@/^1mm/|*\dir{>} "g";"b"
  \ar@{-}@/_1mm/|*\dir{<} "g";"b"
  \ar@{-}@/^1mm/|*\dir{>} "e";"g"
  \ar@{-}@/_1mm/|*\dir{<} "e";"g"
  \ar@{-}@/^1mm/|*\dir{>} "d";"f"
  \ar@{-}@/_1mm/|*\dir{<} "d";"f"
\end{xy}  \label{quiver2}
\end{equation}

It is also clear from (\ref{e:wr4}) that the naive generalization of the monodromy collection (\ref{e:colEn}) cannot be fractional. There is a simple explanation of why the $n=3$ case is singled out. Intuitively there are two types of $\P^1$'s in the $A_{n-1}$ chain of the resolution: the two $\P^1$'s at the end of the chain are different from those in the middle in that they intersect only one other $\P^1$, as opposed to two other $\P^1$'s that a middle $\P^1$ intersects. This makes a big difference in the spectral sequence calculation. For $n=3$ these ``middle'' $\P^1$'s are absent, hence the simplification. We will return to this shortly.

\subsubsection{Seiberg duality}

Now that we know that both (\ref{e:colMcKay}) and  (\ref{e:colEn}) give fractional branes, the question of how they are related arises naturally. We turn to answering it in this section using Seiberg duality.

Seiberg duality was originally formulated \cite{Seiberg:1994pq} as a low-energy equivalence between $N = 1$ supersymmetric gauge theories: an $SU(N_c)$  theory with $N_f$ fundamental flavors  and no superpotential, and an $SU(N_f -  N_c)$ theory with $N_f$ fundamental  magnetic  flavors with a superpotential containing mesons. The duality says that both flow to the same point in the infrared. This was shown to arise as a consequence of an $N = 2  $ duality \cite{Argyres:1996eh}. From our point of view the Berenstein-Douglas \cite{Berenstein:2002fi} extension of Seiberg duality is the relevant one. 

The Berenstein-Douglas formulation of Seiberg duality has a natural stratification. In its simplest form it amounts to a base change for the branes. Since the new basis usually involves anti-branes in the language of the old basis, this change is most naturally done in the derived category of coherent sheaves, rather than sheaves alone. Therefore in this form  Seiberg duality is an autoequivalence of the derived category of coherent sheaves, which by Orlov's theorem  (Theorem \ref{thm:orlov}) is a Fourier-Mukai functor. 

The most general form of Seiberg duality arises when the {\em $t$-structure} of the derived category is changed. This is usually achieved by the use of tilting complexes \cite{Berenstein:2002fi}. What makes this possible is the underlying fact that there are different abelian categories with equivalent derived categories. 

Thus, in general, the difference between two collections of fractional branes can only be partially attributed to a choice of basepoint, since tiltings are more general than auto-equivalences. The McKay collection, although not explicitly, but inherently is associated to the vicinity of the orbifold point. The collection obtained by monodromies explicitly involved the choice of a basepoint for the loops in the moduli space, and this basepoint was in the vicinity of the  large volume point. Therefore it is reasonable to expect that the two collections differ only by a change in basepoint.

Changing the basepoint amounts to conjugating the branes, and as we said earlier is an autoequivalence of the derived category. This gives a Parseval-type equality for the $\Ext$-groups, which leads to the same quiver. This is in line with the fact that both collections gave the same $\C^2/\Z_3$ quiver.		

Indeed, after some educated guesswork, one finds that the two collections (\ref{e:colMcKay}) and  (\ref{e:colEn}) are related by monodromy around a point in moduli space where the brane wrapping both $C_3$ and $C_4$ once is becoming massless.\footnote{The existence of such a point has been established in \cite{DelaOssa:2001xk}.} 
In other words, by Conjecture~\ref{conj:a} the two collections go into each other under the action of $\ms{T}_{\O_{C_3+C_4}}$. First, Eq.~(\ref{e:bpvdw11}) shows that
\begin{equation}
\ms{T}_{\O_{C_3+C_4}}(k_*\O_{C_3+C_4}[1])=\ms{T}_{\O_{C_3+C_4}}(k_*\O_{C_3+C_4})[1]=
k_*\O_{C_3+C_4}[1-1]=k_*\O_{C_3+C_4}\,.
\end{equation}
Second, by a similar computation:
\begin{equation}
\ms{T}_{\O_{C_3+C_4}}(i_*\O_{ C_i})=i_*\O_{ C_i}(-1)[1]\,, \quad \mbox{for $i=3,4$}\,.
\end{equation}

What happens to this relationship for $n>3$? Since we know that the collection 
\begin{equation}\label{eq:lk1}
\O_{\sum C_j} \quad \mbox{and} \quad \O_{C_i}(-1)[1]\,,
\end{equation}
is fractional for any $n$, and for $n=3$ it gave under the Seiberg duality $(\ms{T}_{\O_{C_3+C_4}})^{-1}$ the  $n=3$ version of the collection 
\begin{equation}
\O_{\sum C_j}[1] \quad \mbox{and} \quad \O_{C_i}\,,
\end{equation}
one might ask what is the image of (\ref{eq:lk1}) under $(\ms{T}_{\O_{\sum C_j}})^{-1}$. We need to be more careful here, since the inverse functor $(\ms{T}_{\O_{\sum C_j}})^{-1}$ is not of the form (\ref{e:refl}). It is most simply presented in the form (Definition 2.7 of \cite{ST:braid})
\begin{equation}\label{e:refl2}
\ms{T}^{-1}_{\mathsf{A}}(\mathsf{B}) :=\Cone{ \mathsf{B} \longrightarrow 
lin(\R\!\Hom_{\D(X)}(\mathsf{B},\mathsf{A}), \mathsf{A}) }[-1]\,.
\end{equation}
For a complex of vector spaces $b^{\circ}$, and $ \mathsf{A}\in \D(X)$, the $q$th term in the complex $lin(b^{\circ},\mathsf{A})$ is given by
\begin{equation}
lin^q(b^{\circ},\mathsf{A}) = \prod_{p\in \Z} (A^{q+p})^{\oplus \dim b^p}
\end{equation}

Using this definition, it is easy to check that for an $n$-spherical object $ \mathsf{A}$, one has that  $\ms{T}^{-1}_{\mathsf{A}}(\mathsf{A}) = \mathsf{A}[n-1]$, as expected from Lemma~\ref{lemma:41}. This shows that $\ms{T}^{-1}_ {\O_{\sum C_j}} (\O_{\sum C_j})=\O_{\sum C_j}[1]$.\footnote{One can show by induction using the spectral sequence (\ref{eq:dcx}) that $\O_{\sum C_j}$ is 2-spherical for any $n$.}

Using (\ref{e:refl2}) one obtains
\begin{equation}
\ms{T}^{-1}_ {\O_{\sum C_j}} (\O_{C_i}(-1)[1]) = \left\{ 
\begin{array}{ll}
\O_{\sum_{j\neq i} C_j}  &   \mbox{for  $i = 1$ or $ n-1$}	\\
\O_{C_i}(-1)[1]	 	 &   \mbox{otherwise}	
\end{array}	\right.
\end{equation}
Thus the new collection is
\begin{equation}
\O_{\sum C_j}[1]\,, \quad \O_{\sum_{j>1} C_j}\,, \quad \O_{\sum_{j<n-1} C_j} \quad 
\mbox{and} \quad \O_{C_i}(-1)[1] \quad\mbox{for  $i = 1\ldots n-2$}\,.
\end{equation}
Since we obtained this collection by an autoequivalence from a fractional collection, it is guaranteed to be a fractional collection. This can be checked explicitly.

Finally, let us note that physically the above Seiberg duality acts trivially. This is a consequence of the $N=2$ supersymmetry.\footnote{
For more on Seiberg duality for $N=2$ theories we refer to \cite{Robles-Llana:2004dd}.}

\subsubsection{Quiver $LEGO$}

As we discussed earlier, the resolution of the $\C^2/\Z_n$ singularity introduces $n-1$ $\P^1$'s. In other words the $\C^2/\Z_n$ singularity can be thought of as being produced by the collision of  $n-1$  $\C^2/\Z_2$ singularities. From our earlier results the quiver for  $\C^2/\Z_2$ is simply
\begin{equation}\label{e:colMcKayNv}
 \begin{xy} <1.5mm,0mm>:
  (0,0)*{\bullet}="a"  ,(20,0)*{\bullet}="b",
  ,(0,-4)*{\scriptstyle \O_{C}} ,(20,-4)*{\scriptstyle \O_{C}(-1)[1]}
  ,"b";"b" \ar@(ul,dl)
  ,"a";"a" \ar@(rd,ru)
  \ar@{-}@/^4mm/|*\dir{<}"a";"b"
  \ar@{-}@/_4mm/|*\dir{>}"a";"b"
  \ar@{-}@/^2mm/|*\dir{<} "a";"b"
  \ar@{-}@/_2mm/|*\dir{>} "a";"b"
\end{xy}
\end{equation}
and a fractional collection is given by $\O_{C},\O_{C}(-1)[1]$. 

Eyeballing the collection (\ref{e:colMcKayN}) we see that it contains only the $\O_{C}(-1)[1]$ part of the collection (\ref{e:colMcKayNv}), for every $C_i$. It is natural to assume that the decay of $\O_{C}(-1)[1]$ goes through the channel 
\begin{equation}\label{e:dec6}
\ses{\O_{C}(-1)}{\O_{C}}{\O_p}\,.
\end{equation}
The physics of this decay in the  $\C^2/\Z_2$ case is very simple: in the large volume limit the D3-brane $\O_{p}$, which is space filling and point-like in the internal space, is stable; while at the orbifold point it is marginally unstable under decay  into  $\O_{C}(-1) $ and  $\O_{C}$ \cite{Douglas:2000qw,Aspinwall:2002ke}.

This gives an interesting {\em physical interpretation} of the $\C^2/\Z_n$ result. We have already seen that 
\begin{equation}\label{e:dest2}
\ch(i_*\O_{C}(-1)) =  C\,, \quad
\ch(i_*\O_{C}) =C+p\,.
\end{equation}
Therefore the D5-brane $i_*\O_{C}$ has a D3-brane flux turned on, while  the D5-brane $i_*\O_{C}(-1)$ does not. Now let us look at the chain of  $n-1$ $\P^1$'s in the completely resolved space, and start shrinking say $C_1$. When the volume of $C_1$ reaches $0$, by the above $\C^2/\Z_2$ arguments, the points on $C_1$ become unstable, with decay products $i_*\O_{C_1}$ and $i_*\O_{C_1}(-1)$. 

Next we shrink  $C_2$. This will destabilize the points of $C_2$, which decay into $i_*\O_{C_2}$ and $i_*\O_{C_2}(-1)$, provided that these were stable. Now $C_1$ and $C_2$ intersect in a point, and that point is linearly equivalent to any other point either on $C_1$ or $C_2$. But all these point are destabilized, and hence we expect that so is any brane that carries the charge of such a point. 

This argument can be continued until  all the $C_i$'s have shrunk, at which point (\ref{e:dest2}) would dictate that the $i_*\O_{C_i}$'s have decayed. On the other hand, the $\O_{C}(-1)$'s cannot entirely account for the decay of all the $\O_{C}$'s if we use only the channel (\ref{e:dec6}). This is another reason why $\O_{\sum C_j}$ is needed, and then a repeated use of the octahedral axiom, (\ref{e:bpvdw1}) and (\ref{e:dec6}) give a complete understanding of the decays. This argument therefore suggests that only the  $i_*\O_{C}(-1)$'s are stable at the $\Z_n$ point, and indeed these are the ones that show up in  (\ref{e:colMcKayN}). 

In the above physical argument we used linear equivalence to conclude that any two points on $C_i$ are equivalent. In fact we can deform them into each other. But this is not the case for the singular curve ${\sum C_j}$. This fact can be understood using some technology. First, for an integral scheme $X$ there is an isomorphism between the divisor class group $\operatorname{CaCl} X$, i.e., Weil divisors modulo linear equivalence, and the Picard group $\Pic X$, i.e., Cartier divisors modulo isomorphisms (Prop. II.6.15 of \cite{Hartshorne:}). On the other hand, if $\tilde{C}$ is the normalization of the curve $C$, then \cite{Hartshorne:} provides a short exact sequence connecting $\Pic {C}$ and $\Pic \tilde{C}$:
\begin{equation}\label{e:hart1}
\xymatrix{0 \ar[r] & \bigoplus_{P\in C} \tilde{\O}_P^*/\O_P^* \ar[r] & \Pic C \ar[r] &  \Pic \tilde{C} \ar[r] & 0\,.}
\end{equation}
The normalization of $C = \sum C_j$ is the disjoint union of $n-1$ lines, and therefore $\Pic \tilde{C} = \Z^{\oplus n-1}$. The  Weil divisors on $C = \sum C_j$ are points, and (\ref{e:hart1}) shows that the different points living on the different components are {\em not} linearly equivalent. Furthermore, the divisors supported on the intersection points are also linearly inequivalent. Therefore the curve $C = \sum C_j$ has a rich structure of inequivalent divisors, and it would be interesting to understand what the physical implications of this fact are, e.g., in connection with moduli stabilization.

\subsection{The $\C^2/\Z_n$ quiver from partial resolutions}

\begin{figure}[ht]
\begin{equation}\nonumber
\begin{xy} <1.5cm,0cm>:
{\ar 0;(0,1) }, (0,1.2)*\txt{$v_1$=(0,1)}
,{\ar 0;(2,-1) *+!LD{v_k=(k,1-k)}}
,{\ar 0;(3,-2) *+!LD{v_n=(n,-n+1)}}
,{\ar@{-}@{.>} (0,1);(3,-2) }
\end{xy}
\end{equation}
  \caption{The toric fan for a partial resolution of the $\C^2/\Z_n$ singularity.}
  \label{f:fan1}
\end{figure}
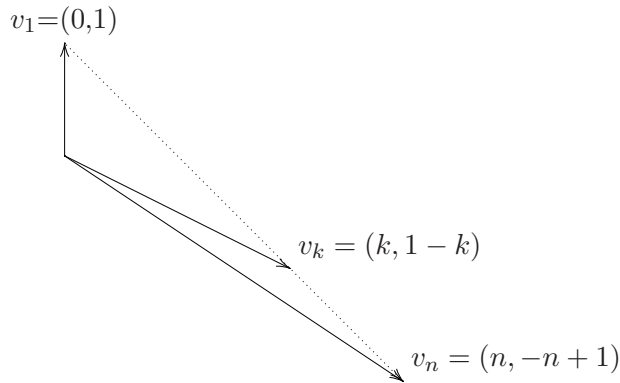

The partial resolutions of the $\C^2/\Z_n$ singularity form a partially ordered set. The simplest partial resolutions involve blowing up {\em only one} of the $n-1$ exceptional divisors. This is particularly easy to do torically. We sketched the general situation in Fig.~\ref{f:fan1}.

The McKay correspondence  gives an equivalence between quiver representations and sheaves on the resolved space, but it glosses over the partial resolutions. One can fill in the gap, by recasting it slightly into the language of stacks. First recall that there is an equivalence of categories between $\C^n/G$ quiver representations and coherent sheaves on the quotient stack  $[\C^n/G]$. Therefore the McKay correspondence reads as
\begin{equation}
\D([\C^n/G])\cong \D(\mbox{crepant resolution of $\C^n/G$})\,.
\end{equation}

Kawamata generalized the above statement, and for $G$ abelian he proved that \cite{Kawamata:GMcKay}:
\begin{equation}
\D([\C^n/G])\cong \D(\mbox{{\bf partial} crepant {\bf stacky} resolution})\cong \D(\mbox{crepant resolution})
\end{equation}
where in the middle one has to consider the partially resolved space as a stack. 

Therefore it makes sense to talk about fractional branes on the partially resolved space, and ask what they are. The strategy of this section is to use an appropriate set of objects on the exceptional divisor of the resolution to model the fractional branes. This strategy was successfully deployed in \cite{Herzog:2005sy} as well.

All the lattice points corresponding to the crepant partial resolutions lie on the line $x+y=1$, and are of the form $v_k=(k,1-k)$, where $1\leq k \leq n-1$. These points are equidistant, and therefore the  star of the associated toric divisor $D_k$ is given by the following 1-dimensional fan: 
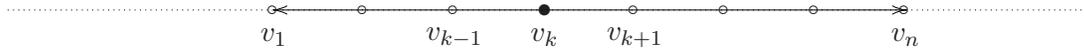
\begin{equation}\nonumber
\begin{xy} <1.2cm,0cm>:
,{\ar@{<->} (-3,0);(4,0)}
,(0,0)*{\bullet}	,(-0,-0.3)*{v_k}
,(-1,0)*{\comp}	,(-1,-0.3)*{v_{k-1}}	
,(-2,0)*{\comp} 	
,(-3,0)*{\comp}	,(-3,-0.3)*{v_{1}}		
,(1,0)*{\comp}	,(1,-0.3)*{v_{k+1}}		
,(2,0)*{\comp}
,(3,0)*{\comp}
,(4,0)*{\comp}	,(4,-0.3)*{v_{n}}	
,{\ar@{-}@{.} (-6,0);(6,0) }
\end{xy}
\end{equation}

One immediately recognizes this fan as corresponding to the weighted projective line $\P^1(k,n-k)$. Without loss of generality we can assume that $k<n-k$. As a variety, or scheme, $\P^1(k,n-k)$ is isomorphic to $\P^1$. The origin of this ``smoothing-out'' is identical to the one that underlies the isomorphism ${\mathbb C} / {\mathbb Z}_n \cong {\mathbb C}$. Algebraically this isomorphism is
 in fact a trivial statement: $\Spec \C[x^n] \cong \Spec \C[x]$. 

In order for the weighted projective line $\P^1(k,n-k)$ to be able to capture the fact that it provides a  partial resolution for the $\C^2/\Z_n$ singularity, we have to retain more information than it's scheme structure. In fact, the toric fan contains this data. We choose to retain this extra embedding information by using the language of stacks. 

We can consider the stack ${\mathbf P}^1(a,b)$ from two different points of view: as a toric stack \cite{BorisovDM}, or as a quotient stack \cite{Auroux,Alberto}. We find it very convenient to work with the latter description.

The stack $\mathbf{P}^1(k,n-k)$ has a full and strong exceptional collection of length $n$ \cite{Auroux,Alberto}:
\begin{equation}
\O,\; \O(1)\; \ldots\; \O(n-1)\,.
\end{equation}
The mutation-theoretic dual of this exceptional collection was throughly investigated in \cite{Alberto}.\footnote{For the definition and properties of mutations see, e.g., \cite{Rudakov:Book}.} 
In particular, Proposition 2.5.11 of \cite{Alberto} states that the mutation-theoretic left dual of the collection $\O, \ldots , \O(n-1)$  is given by the full exceptional sequence 
\begin{equation}
{\c M}_{(1-n)}[1 - n]\,,\; {\c M}_{(2-n)}[2 - n]\,,\; \ldots \,,\; {\c M}_{(-1)}[-1]\,,\; {\c M}_{(0)}\,.
\end{equation}

In order to explain the previous expression we need to introduce some notation. Let $I \subseteq \{1, 2\}$ be a subset, and consider the  weighted projective line $\mathbf{P}^1(w_1,w_2)$. Then  $\# I$ will denote the number of elements in  $I$, while $|w_I|=\sum_{i\in I}w_i$. In this notation, for $0\leq l <n$, the complex ${\c M}_{(-l)}$ is defined as  a subcomplex of the Koszul complex  $\c K$ twisted by $\O(-l)$ \cite{Alberto}, with $j$th term given by:
\begin{equation}
{\c M}^j_{(-l)} := \bigoplus_{\#I=-j,|w_I|=l} \O(l - |w_I |) \subseteq \bigoplus_{\#I=-j} \O(l - |w_I |) = \c K^j(-l)\,.
\end{equation}
In other words ${\c M}_{(l)}$ has non-zero components only in non-positive degrees.

For  the stack $\mathbf{P}^1(k,n-k)$ the explicit expressions for the ${\c M}_{(l)}$'s are easy to write down:
\begin{equation}\label{eq:albi12}
{\c M}_{(-l)} = \left\{ 
\begin{array}{rcl}
\xymatrix@1{\poso{\O(l)}} & 	 						for  & 0\leq l<k \\
\xymatrix@1{\O(l-k)\ar[r]& 		  \poso{\O(l)}} &	 		for  & k\leq l<n-k	\\
\xymatrix@1{\O(l-k)\oplus \O(l-n+k)\ar[r]&  \poso{\O(l)}}	  &   for  & l\geq n-k	
\end{array}	\right.
\end{equation}

For brevity let us denote the stack $\mathbf{P}^1(k,n-k)$ by ${\cal Y}_k$. Similarly, the partially resolved quotient stack $\mathrm{Bl}_k [\C^2/\Z_n]$, with exceptional divisor $D_k$, is denoted by  ${\cal X}_k$. Let 
$$i\!: {\cal Y}_k=\mathbf{P}^1(k,n-k) \longrightarrow  {\cal X}_k=\mathrm{Bl}_k [\C^2/\Z_n]$$
denote the embedding morphism of stacks. 
\begin{prop}
For $n$ and $k$ relatively prime, the pushed-forward complexes 
$$i_*\c M_{(1-n)}\,,\;  i_*\c M_{(2-n)}\,,\; \ldots \,,\;  i_*\c M_{(-1)}\,,\;  i_*\c M_{(0)}$$ 
provide a model for the $\C^2/\Z_n$ fractional branes.
\end{prop}

\begin{proof}
First we show that the $\Ext$-quiver of the collection $i_*\c M_{(1-n)},\ldots , i_*\c M_{(-1)},i_*\c M_{(0)}$ is the $\C^2/\Z_n$ quiver. To evaluate the $\Ext$-groups we use the stacky version of the spectral sequence (\ref{SS1}), as presented for example in \cite{Eric:DC2}, and adapted to our case $i\!: {\cal Y}=\mathbf{P}^1(k,n-k) \hookrightarrow  {\cal X}=\mathrm{Bl}_k [\C^2/\Z_n]$.\footnote{For simplicity we dropped the subscript $k$.}
For ${\cal E}$ and $ \cal F$ two objects in the bounded derived category of the stack ${\cal Y}=\mathbf{P}^1(k,n-k)$ the spectral sequence reads
\begin{equation}\label{SS1a}
E_2^{p,q}=
\Ext^p_{\c Y}( {\cal E} ,\, {\cal F} \otimes \Lambda^q N_{\c Y/\c X}) \: \Longrightarrow \:
\Ext^{p+q}_{\c X}\left( i_* {\cal E}, i_* {\cal F} \right)\,.
\end{equation}
Since $N_{\c Y/\c X}=K_{\c Y}$ has rank one,\footnote{We used the fact that $K_{\c X}$ is trivial. } the spectral sequence degenerates at $E_2$, and we have that 
\begin{equation}
\Ext^1_{\c X}\left( i_* {\cal E}, i_* {\cal F} \right)=\Ext^1_{\c Y}( {\cal E} ,\, {\cal F})\oplus \Ext^0_{\c Y}( {\cal E} ,\, {\cal F}\otimes K_{\c Y})\,,
\end{equation}
Serre duality then gives $\Ext^0_{\c Y}( {\cal E} ,\, {\cal F}\otimes K_{\c Y})=\Ext^1_{\c Y}( {\cal F} ,\, {\cal E})^\vee$, and therefore
\begin{equation}\label{SS1b}
\Ext^1_{\c X}\left( i_* {\cal E}, i_* {\cal F} \right)=\Ext^1_{\c Y}( {\cal E} ,\, {\cal F})\oplus \Ext^1_{\c Y}( {\cal F} ,\, {\cal E})^\vee\,.
\end{equation}
This shows that $\Ext^1_{\c X}$ is automatically symmetric, and the resulting quiver has bidirectional arrows.

The  $\Ext^1_{\c X}(\c M( l), \c M( j))$ groups are easily computed using Lemma~2.5.12 of \cite{Alberto}:
\begin{equation}
\dim \Ext^k_{\c X}(\c M( l),\c M( j)) =  \#\{J \subseteq \{1, 2\} \, |\, \#J = k, |w_J | = j -l\}\,.
\end{equation}

This observation reduces the problem of determining the $\Ext$-quiver to a purely graph theoretic one, which is easy to solve. Let the $n$ objects $\c M_{(1-n)}, \c M_{(2-n)},\ldots , \c M_{(-1)},\c M_{(0)}$ be the nodes of a graph. For every non-zero $\Ext^1_{\c Y}(\c M_{(-i)},\c M_{(-j)})$ we put a directed arrow. For the first $k$ nodes, i.e.,  $1\leq l < k$, there are two outgoing arrows (one to node $l+k$ and another one to node $l+n-k$). Similarly, for $k\leq l \leq n-k$ there is one incoming and one outgoing arrow. Finally, for $l> n-k$ there are two incoming arrows. Therefore we have a graph with two arrows originating or ending at every node. 

Going from the collection $\c M_{(1-n)},\ldots , \c M_{(-1)},\c M_{(0)}$ to the collection $i_*\c M_{(1-n)}$,  $\ldots$,  $ i_*\c M_{(-1)}$,  $i_*\c M_{(0)}$ by virtue of  Eq.~(\ref{SS1b}) simply makes every arrow bidirectional. So the orientations of the arrows can be dropped, and they become paths.  

Now we have a graph such that every node is visited by two paths. After reordering this is precisely the $\C^2/\Z_n$ quiver, provided that we can show that the graph is connected. This is where the $\gcd(n,k)=1$ condition comes in, which is equivalent to the condition $\gcd(n-k,k)=1$. The Euclidian algorithm then guarantees the existence of two integers $a$ and $b$ such that 
$$a(n-k)+b k=1.$$
Since we have only links of length $n-k$ and $k$ the above equation shows that there is a path connecting any two neighboring nodes, where the path in question has $a$ links of length $n-k$ and $b$ links of length $n$. This establishes that the $\Ext$-quiver of the collection $i_*\c M_{(1-n)},\ldots , i_*\c M_{(-1)},i_*\c M_{(0)}$ is connected, and hence it is the $\C^2/\Z_n$ quiver.

The next step is to show that the $i_*\c M_{(l)}$'s  indeed ``add up'' to the D3-brane. For this we need to compute $\sum_{i=0}^{1-n} \ch i_*\c M_{(l)} $. We use the stacky version \cite{Toen:GRR} of the Grothendieck-Riemann-Roch theorem (GRR):
\begin{equation}
\sum_{l=0}^{1-n} \ch(i_*\c M_{(l)})\td(\c X)=\sum_{l=0}^{1-n} i_*(\ch \c M_{(l)}\td \c Y )= i_*(\left(\sum_{l=0}^{1-n}\ch \c M_{(l)} \right)\td \c Y )\,.
\end{equation}
The Chern characters $\ch \c M_{(l)} $ are straightforward to compute from (\ref{eq:albi12}), in terms of the class of a point on ${\cal Y}=\mathbf{P}^1(k,n-k)$:
\begin{equation}
\begin{split}
\sum_{l=0}^{1-n}\ch \c M_{(l)} &=\sum_{l=0}^{k-1}e^{lH}+\sum_{l=k}^{n-k-1}e^{lH}-e^{(l-k)H}
+\sum_{l=n-k}^{n-1}e^{lH}-e^{(l-k)H}-e^{(l-n+k)H}\\
 &=\sum_{l=0}^{n-1}e^{lH}-\sum_{l=0}^{k-1}e^{lH}-\sum_{l=0}^{n-k-1}e^{lH}=k(n-k)H
\end{split}
\end{equation}
In the toric description of ${\cal Y}=\mathbf{P}^1(k,n-k)$ \cite{BorisovDM} it is clear that $k(n-k)H$ is the Chern character of a ``non-singular'' point on the stack ${\cal Y}=\mathbf{P}^1(k,n-k)$.\footnote{This guarantees that we need not deal with the subtleties of the inertia stack.} The GRR theorem then shows that $\sum_{i=0}^{1-m} \ch i_*\c M_{(l)} $ equals the Chern character of a non-singular point on $\c X$, which is what we wanted to prove.
\end{proof}

We believe that a similar result holds even without the technical condition $\gcd(n,k)=1$. More precisely, if $\gcd(n-k,k)=d$ and $n-k=n_1 d$ and  $k=n_2 d$, then the stack  $\mathbf{P}^1(k,n-k)$ is a $\Z_d$ quotient of $\mathbf{P}^1(n_1,n_2)$. Let $\c D$ be a $d$-torsion divisor on $\mathbf{P}^1(k,n-k)$, and  $\c H$ be the hyperplane-divisor. Then one would have to consider the mutation-theoretic dual of the exceptional collection  
$$\O, \ldots , \O(n\c H), \O(\c D), \ldots , \O(\c D+n\c H), \ldots ,\O((d-1)\c D), \ldots , \O((d-1)\c D+n\c H)\,. $$  
But proving this is beyond the scope of the present paper. Instead we just note that a similar problem was treated in \cite{Herzog:2005sy}. 

\subsection{BPS algebras}

Ten years ago, while computing threshold corrections in $\c N = 2$ heterotic compactifications, Harvey and Moore observed that these were closely related to product formulas in generalized Kac-Moody (GKM) algebras. Inspired by the heterotic/IIA duality they introduced {\em the algebra of BPS states} \cite{Harvey:1995fq,Harvey:1996gc} to explain the appearance of the GKM algebras. The product structure was defined in terms of on integral on the correspondence variety of certain  moduli spaces of sheaves.

At that time D-branes were thought of as sheaves with charge  valued in K-theory \cite{Minasian:1997mm,Witten:1998cd}. Currently we have a refinement of this picture, with D-branes as objects in $\D(X)$, while the charge is determined by the natural map $\D(X)\to \K(\D(X))=\K(X)$. Therefore one needs to revisit the way the algebra of BPS states is defined. 

An algebraic model for the  algebra of BPS states could be provided by  Ringel-Hall algebras. For an abelian category $\c A$ in which all the $\Ext^1$ groups are finite, Ringel \cite{Ringel} defines an algebra $R(\c A)$, which is the free abelian group on the isomorphism classes of $\c A$, endowed with a multiplication whose structure constants are suitably normalized $\Ext^1$'s. Unfortunately it seems hard to extend this construction to a triangulated category, such as  $\D (\c A)$. Mimicking  the Hall algebra construction, with exact triangles replacing exact sequences, fails to give an associative multiplication \cite{Kapranov:Heisenberg}. 

This suggests that one should perhaps define the BPS algebra of D-branes in terms of abelian categories. This seems possible throughout the moduli space $\c M_{\c K}$ of complexified Kahler forms. Douglas \cite{Douglas:2000gi} proposed that $\c M_{\c K}$ can be covered by open subsets $U_i$, and for each $U_i$ there is an abelian category $\c A_i$, such that $\D(\c A_i)\iso \D(X)$. This was proven recently in a different form for the local \CY\ $\O_{\P^2}(-3)$ by Bridgeland \cite{BridgelandC3Z3}. 

Bridgeland proves that for every point in an open subset $Stab^0(X) \subset  Stab(X)$ of the space of $\pi$-stability conditions there exists a bounded $t$-structure, and that the heart of this $t$-structure, which is an abelian category, is invariant for an open subset of $Stab^0(X)$. Therefore the natural proposal is to take the algebra of BPS  states to be the Ringel-Hall algebra of the {\em $\pi$-stable objects}.  

On the other hand, defining the BPS algebra of D-branes in terms of abelian categories seems aesthetically unsatisfactory. 
Another approach would be to consider all BPS D-branes, not only the stable ones. 
As we move in the Kahler moduli space $\c M_{\c K}$, the collection of stable objects changes, and so does our proposed BPS algebra. The most extreme case of this is Seiberg-Witten theory \cite{SW:I}. For simplicity let us focus on the pure $\c N=2$ case with gauge group $\SU(2)$. Geometric engineering tells us how to obtain this theory from string theory, and the BPS states of the field theory become a subset of those of the string theory. In the derived category context this map was worked out in \cite{en:Paul}. The spectrum of stable objects is particularly simple: there is an almost circle-shaped region delimited by two lines of marginal stability. Outside of it there are infinitely many BPS states, but inside there are only two: the monopole and the dyon. Therefore the BPS algebra jumps from an infinite dimensional one to become finite dimensional.

At this point very little is know about the algebras of BPS states. Ringel originally proved the following: let $\Delta$ be a simply laced Dynkin diagram $\Delta$, with $\vec{\Delta}$ the associated quiver, $\mathfrak{g}_{\Delta}$ the corresponding Lie algebra, and let $\c A(\vec{\Delta})$ be the abelian category of finite-dimensional representations of $\vec{\Delta}$. Then $R(\c A(\vec{\Delta}))$ is the positive part (nilpotent subalgebra) of the enveloping Hopf algebra $\U(\mathfrak{g}_{\Delta})$. Working over the finite field $\F_q$, $R(\c A(\vec{\Delta}))$ is  the positive part of the quantum group $\U_q(\mathfrak{g}_{\Delta}) $. The same result was proven in  \cite{Kapranov:Vasserot} using the derived category of the resolution of  $\C^2/G_{\Delta}$, where $G_\Delta$ is the finite subgroup of $\SL(2,\C)$ corresponding to $\Delta$. Ringel-Hall algebras were also investigated recently for elliptic curves \cite{Burban:Schiffmann}.

It would be even more interesting if one could associate an algebra $DR(X)$ to $\D(X)$ directly, without resorting to its  abelian hearts. Moreover, one would like to see the $R(\c A_i)$ heart-algebras as subalgebras in $DR(X)$. It was suggested that $DR(\c A(\vec{\Delta}))$ might yield the whole quantum group \cite{Kapranov:Heisenberg}. $DR(X)$ would have interesting physical content, as it would integrate both perturbative and non-perturbative information, and would be inherently characteristic of the D-branes throughout the entire moduli space. 

Progress in this direction has recently been made by Toen \cite{Toen:Hall}, who defines an associative multiplication on the
rational vector space generated by the isomorphism classes of a triangulated category $\c T$, where $\c T$ is the perfect derived category $per( T)$ of a proper dg-category $ T$. The resulting $\Q$-algebra is the derived Hall algebra $\cal{DH}(T)$. The multiplication involves all  the $\Ext^i$'s, not only $\Ext^1$, and therefore is more natural.

Toen proves that if $\c A$ is the heart of a non-degenerate t-structure, then the Ringel-Hall algebra $R(\c A)$ is a subalgebra of $\cal{DH}(T)$. Unfortunately for us, the relationship between $\cal{DH}(\D(\c A(\vec{\Delta})))$ and the quantum group $\U_q(\mathfrak{g}_{\Delta})$ remains to be investigated, and an explicit example is yet to be worked out.

\section*{Acknowledgments}

It is a pleasure to thank Alastair Craw, Alberto Canonaco, Emanuel Diaconescu, Mike Douglas, Chris Herzog, Greg Moore,  Tony Pantev, Ronen Plesser, Martin Rocek, Seiji Terashima and Eric Sharpe for useful conversations.  I am especially indebted to Paul Aspinwall, from whom I learned most of the subject, for discussions at early stages of this project. I would also like to thank the 2005 Simons Workshop on Mathematics and Physics, and the 2005 Summer Institute in Algebraic Geometry at the University of Washington, for providing a stimulating environment where part of this work was done.
The author was supported in part by the DOE grant DE-FG02-96ER40949.

\begin{appendix}

\section{Some useful spectral sequences}\label{s:ss} 

In the bulk of the paper we make extensive use of spectral sequences. This is a well known device in algebraic geometry, but so far had a limited appearance in the physics literature. The three spectral sequences we use have got an extensive recent treatment \cite{Eric:DC1}. Let us state them one by one, following the presentation of \cite{Eric:DC1}, and at the same time rephrasing their results in terms of $\Ext$ groups.

The simplest case concerns a {\em smooth} subvariety $S$ of a smooth variety $X$. Let $i\!: S \hookrightarrow X$ be the embedding, and $N_{S/X}$ the normal bundle of $S$ in $X$. Then for two vector bundles ${\cal E}$ and ${\cal F}$, or more precisely locally free sheaves on $S$, we have the first  spectral sequence:
\begin{equation}\label{SS1}
E_2^{p,q}=
\Ext^p_S( {\cal E} ,\, {\cal F} \otimes \Lambda^q N_{S/X}) \: \Longrightarrow \:
{\Ext}^{p+q}_X\left( i_* {\cal E}, i_* {\cal F} \right)
\end{equation}
where $\Lambda^q$ denotes the $q$th exterior power. 

A more general case is when you are given two nested embeddings: $j\!: T \hookrightarrow S$ and $i\!: S \hookrightarrow X$,  a  vector bundle  $\c F$ on $T$,  and a vector bundle $\c E$ on $S$. Then we have the spectral sequence:
\begin{equation}\label{SS2}
E_2^{p,q}= 
\Ext^p_T ({\cal E}|_T ,\,  {\cal F} \otimes \Lambda^q N_{S/X}|_T )
\: \Longrightarrow \: \Ext^{p+q}_X\left( i_* {\cal E}, j_* {\cal F} \right)
\end{equation}
The symbol $|_T$ means restriction to $T$.

The final and most general case deals with two subvarieties $T$ and $S$ of $X$. Now the embeddings are $i\!: S \hookrightarrow X$ and $j\!: T \hookrightarrow X$. Once again $\c F$ is a vector bundle on $T$,  and $\c E$ is a vector bundle  on   $S$. The spectral sequence is:
\begin{equation}\label{SS3}
E_2^{p,q} = \Ext^p_{S\cap T}({\cal E}|_{S \cap T},\,  
{\cal F}|_{S \cap T} \otimes \Lambda^{q-m} \tilde{N} \otimes \Lambda^{top} N_{S \cap T / T} )\Longrightarrow
{\Ext}^{p+q}_X \left( i_* {\cal E}, j_* {\cal F} \right)
\end{equation}
where $ \tilde{N}= TX|_{S \cap T}/(TS|_{S \cap T} \oplus TT|_{S \cap T})$ is a quotient of tangent bundles, while $m=\rk N_{S \cap T / T}$.

Although these spectral sequences were derived for sheaves, they extend to the derived category. It is also clear that (\ref{SS1}) is a particular case of (\ref{SS2}), which in turn is a particular case of (\ref{SS3}). 

\section{A simple spectral sequence}\label{a:1}

In this appendix we derive a spectral sequence that is used extensively in the paper. Our derivation follows ideas from the Appendix of \cite{Eric:DC1}. 

Let $X$ be a smooth algebraic variety. Consider two divisors  $C$ and $D$ on $X$, and the embedding maps: $i \! : C+D  \hookrightarrow X$ and $j \! : C \hookrightarrow X$. Our task is to compute $\Ext_X^i(i_*\O_{C+D}, j_*\c F)$ for a coherent sheaf $\c F$ on $C$. It is worth pointing out that the divisor $C+D$ is singular. 

We start with the short exact sequence of sheaves on $X$
\begin{equation}
\xymatrix{0 \ar[r] & \O_X(-C-D)  \ar[r]^-s & \O_X \ar[r] & i_*\O_{C+D} \ar[r] & 0\,.}
\end{equation}
This is a locally free, and therefore projective resolution for $i_*\O_{C+D}$. Now we apply the left exact contravariant functor $\sHom_X(-, j_*\c F)$
\begin{equation}
\xymatrix{0 \ar[r] & \sHom_X(i_*\O_{C+D}, j_*\c F) \ar[r]^-{s^\sharp} &  \sHom_X( \O_X, j_*\c F)\ar[r] & 
\sHom_X( \O_X(-C-D) , j_*\c F)  }
\end{equation}
The $\Ext$ groups are given by the homology of this complex.
Since ${\rm supp} (j_*\c F) \subset C$, and $s$ vanishes on $C+D$, it follows that $s^\sharp=0$. Therefore
\begin{equation}
\begin{split}
&\sExt^0(i_*\O_{C+D}, j_*\c F) =  \sHom_X( \O_X, j_*\c F) 		= j_*\c F \\
&\sExt^1(i_*\O_{C+D}, j_*\c F) =  \sHom_X( \O_X(-C-D) , j_*\c F)  = j_*\c F (C^2+CD)\,.
\end{split}
\end{equation}

At this point we can use the local to global spectral sequence on $X$
\begin{equation}
E_2^{p,q} = \H^p (X, \sExt^q_X(\c A , \c B) ) \Longrightarrow \Ext_X^{p+q} (\c A, \c B)\,,
\end{equation}
and the fact that $\H^p (X, j_*\c A) =  \H^p (C, A)$ to conclude that the spectral sequence with the following $E_2^{p,q}$ term
\begin{equation}\label{eq:dcx}
\begin{xy}
\xymatrix@C=10mm{
  & 0& \\
   & \H^p (C,  \c F (C^2\!+\!CD))\\
   & \H^p (C,  \c F )}
\save="x"!LD+<-3mm,0pt>;"x"!RD+<0pt,0pt>**\dir{-}?>*\dir{>}\restore
\save="x"!LD+<0pt,-3mm>;"x"!LU+<0pt,-2mm>**\dir{-}?>*\dir{>}\restore
\save!RD+<0mm,-4mm>*{p}\restore
\save!LU+<-3mm,-5mm>*{q}\restore
\end{xy}
\end{equation}
converges to $\Ext_X^{p+q}(i_*\O_{C+D}, j_*\c F)$.

\end{appendix}



\end{document}